\documentclass[preprint2]{proto}
\pdfoutput=1
\usepackage{times}
\usepackage{xcolor}
\usepackage{graphicx}
\usepackage{wasysym}
\usepackage{float}
\usepackage{dblfloatfix}
\usepackage{amssymb}
\usepackage[T1]{fontenc}

\graphicspath{{./}{Figures/}}

\begin{document}

\title{\textbf{\LARGE Interstellar Heritage and the Birth Environment of the Solar System}}

\author {\textbf{\large Edwin Bergin}}
\affil{\small\em University of Michigan}

\author {\textbf{\large Conel Alexander}}
\affil{\small\em Carnegie Institution for Science}

\author {\textbf{\large Maria Drozdovskaya}}
\affil{\small\em Universit\"{a}t Bern}

\author {\textbf{\large Matthieu Gounelle}}
\affil{\small\em Mus\'eum national d'Histoire naturelle}

\author {\textbf{\large Susanne Pfalzner}}
\affil{\small\em J\"{u}lich Supercomputing Center, Max-Planck-Institut f\"{u}r Radioastronomie}

\begin{abstract}

\begin{list}{ } {\rightmargin 1in}
\baselineskip = 11pt
\parindent=1pc
{\small 
In this chapter, we explore the origins of cometary material and discuss the clues cometary composition provides in the context of the origin of our solar system.
  The review focuses on both cometary refractory and volatile materials, which jointly provide crucial information about the processes that shaped the solar system into what it is today. Both areas have significantly advanced over the past decade. We also view comets more broadly and discuss compositions considering laboratory studies of cometary materials, including interplanetary dust particles and meteoritic material that are potential cometary samples, along with meteorites, and {\em in situ}/remote studies of cometary comae. In our review, we focus on key areas from elemental/molecular compositions,  isotopic ratios, carbonaceous and silicate refractories, short-lived radionuclides, and solar system dynamics that can be used as probes of the solar birth environment. We synthesize this data that points towards the birth of our solar system in a clustered star-forming environment.
\\~\\~\\~}
\end{list}
\end{abstract}  

\section{\textbf{INTRODUCTION}}
\label{sec:intro}

Comets are generally posited as providing samples of the ``most'' primordial material in the solar system.   In this context, it is important to understand what primordial means.  Generically, the limit is when our solar system becomes isolated from the interstellar medium (ISM).  Do the comets bear an imprint of the physicochemical processes active during planet formation or do they preserve a history of the stages that preceded the birth of the Sun? There is evidence in the cometary record of both aspects with  processed solids, combined with potentially unaltered icy and refractory material.\footnote{In general, astronomical literature makes a distinction between refractory, which are effectively the ``rocky'' solids, and volatile material, which are found as ice and gas.  The distinction between these components is the condensation/sublimation temperature, which is substantially higher for refractories (e.g., silicate minerals and macro-molecular organics) than for volatiles (e.g., H$_2$, H$_2$O, and CO).}
This is of interest as the composition of primordial material might therefore relate to the birth environment.  Our goal is to understand the full context of this perspective and leave the question of the chemistry that is active during the phase of planet formation to the chapter by Aikawa (et al.).  For the remainder of this section, we will give a general description of galactic star formation and its stages, delineate some key cometary properties, discuss their meteoritic analogs from the inner ($<$ 5~au) solar system, and discuss explorations of cometary material in the laboratories on Earth. In \S~2, we describe the general evolution from the diffuse ISM to the molecular cloud and explore which aspects of the cometary record reflect this stage, while \S~3 discusses the protostar and protostellar disk stages and their relevance to cometary composition.   In \S~4, we summarize the current understanding of the birth environment of the solar system as traced by comets. In relevant subsections we directly summarize the available constraints.

\subsection{Star, planet formation, and the birth environment}

\begin{figure*}[ht!]
\begin{center}
\includegraphics[width=15cm]{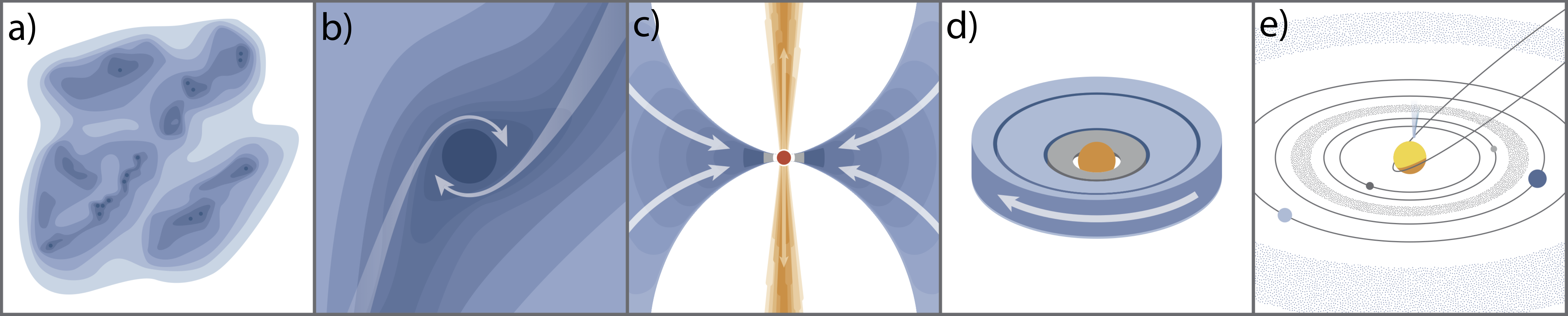}
\caption{Schematic of star and planet formation taken from \citet{Oberg21} with image credit to K. Peek.  Stars and planets are born in (a) molecular clouds which have filamentary substructure and are comprised of gas and dust. (b) Rotating centrally concentrated dense cores are embedded within the filaments; these are labelled as pre-stellar cores.  (c) The rotating pre-stellar core collapses leading to the formation  of a protostar (Class 0 and I), surrounded by a forming {\em protostellar} disk which is still accreting material from the natal envelope.  The presence of accretion onto the forming star produces a bipolar outflow.  (d) Overtime the stellar wind and outflow dissipates the surrounding envelope leaving an exposed gas- and dust-rich {\em protoplanetary} disk (labelled as Class II).  Giant planets and the seeds of terrestrial worlds form in this stage with comets forming in regions beyond the disk water ice sublimation front.  (e) After gas disk dissipation  and the final stages of terrestrial planet formation, a planetary system emerges.   }
\label{fig:f1}
\end{center}
\end{figure*}

 In Fig.~\ref{fig:f1} we provide a cartoon of the stages of star and planet formation and here provide a brief description of star and planet formation with specifics provided in subsequent sections.  Stars are born in clouds of gas, that is predominantly molecular in composition, and small dust particles with an average size of 0.1~$\mu$m.  The gas is H$_2$ rich with dilute, but measurable, quantities of other molecular species (e.g., CO, H$_2$O ice and vapor, CO$_2$, organics, etc.), while the dust is both silicate and carbonaceous in composition \citep{Draine03}.  Dense (n$_H > 10^4$~cm$^{-3}$) gas in molecular clouds is concentrated in filamentary structure.  More centrally concentrated ``pre-stellar cores'' are embedded within the filaments in a phase that lasts for $\sim$1~Myr \citep{Andre14}.  
These cores condense and collapse to form protostars.  Astronomers designate protostars as either Class 0 or Class I.  Class 0 are younger protostars where a young {\em protostellar} disk forms surrounded by a dense collapsing envelope \citep{Andre00}.  Young stars actively accrete from this envelope, perhaps episodically \citep{Audard14}.  They also generate well-characterized bipolar outflows and outflows, along with disk winds. These collectively ablate the surrounding envelope \citep{Arce07}.  This leads to a disk with a much-reduced natal envelope, thereby ensuring the Class I stage.  The lifetime of the protostellar stage is short, lasting only of order 100,000 years \citep{Kristensen18}. 
   As time proceeds, the envelope dissipates and the disk-star system is exposed to the interstellar radiation field - the so-called protoplanetary disk or Class II stage, which has a half-life time estimate of $\sim 2$~Myr.  Within the protostellar and protoplanetary disk dust grains coagulate to $\sim$mm/cm sizes (called pebbles) and concurrently sink to the dust-rich disk midplane where, depending on their size, they are subjected to differential forces that leads to inwards pebble drift \citep{Andrews:2020}.   In the solar system, comets themselves formed either in the protostellar or protoplanetary phase in the outer reaches of the disk in proximity or beyond the location of gas and ice giant planets.
 
Statistics of star-forming regions demonstrate that most stars are born in clusters \citep{Lada:2003} and previous analyses of the birth environment of the solar system have suggested that the Sun was born in a cluster \citep[e.g.,][]{Adams:2010, Pfalzner:2020}.   A central facet of clustered star formation is that all massive ($>$1~M$_{\odot}$) stars are born in clusters and, thus, birth in a cluster could mean the nearby presence of one or more massive ($> 1$~M$_\odot$) stars depending on the size of the cluster.  Looking at local star formation, there is a question as to whether there are distinct clustered or distributed modes.  \citet{Bressert10} find a continuum of star formation surface density with a log-normal function and a peak of young stellar object (YSO) surface densities at 22 YSO pc$^{-2}$.   \citet{Megeath22} explored the YSO surface density in eight nearby clouds.  They drive a distinction between clouds with high $> 25$ YSO pc$^{-2}$ (e.g., Ophiuchus, located near massive stars including the Sco OB2 association) and clouds without clusters (e.g., Taurus with no nearby massive stars) with low values, $<$ 10 YSO pc$^{-2}$.  

In this review, we  explore a perspective for the overall chemistry in terms of the presence or absence of a nearby presence of massive star.  This then would place limits on the birth environment in terms of a modestly to very rich cluster of stars.  The alternate would be a small cluster of stars or more distributed star formation such as seen in some nearby star-forming clouds (e.g., Taurus).   {\em For the purposes of this review we will refer to this difference as ``clustered vs. isolation'', where the key dividing line is the presence or absence of a massive star.} A nearby massive star leaves an imprint on the composition of cometary ices and refractories.  If the Sun was born in a cluster with a nearby massive star the external energetic  radiation field would indirectly heat the surrounding molecular cloud material, via reprocessed radiation, and directly expose surfaces to photo-dissociating/ionizing radiation.  This can change the chemistry of cometary volatiles. Winds from the star, or the supernova explosion could seed the forming cloud or core with active radionuclides and newly formed grains.

\subsection{General information about comets}

Comets are the most distant residents of our solar system. The two main reservoirs of cometary bodies are located in the outer sectors of the planetary system of the Sun. The first is the Kuiper belt, including the Scattered disk, that extends at radii between $\sim30-50$~au, beyond the orbit of Neptune, and encompasses objects of various sizes (from smaller than $1$~km to several hundreds of km in diameter), binaries, bodies with rings, and several dwarf planets such as Pluto. The second is the Oort cloud, located at distances of $2,000-200,000$~au and containing a much larger number of bodies (billions in comparison to several $100,000$s in the Kuiper belt), but of smaller dimensions (few km at most). The Kuiper belt is donut-shaped, dynamically stable, and generally gives rise to short-period comets upon small gravitational perturbations from Neptune. On the other hand, the Oort cloud is thought to be spherical and dynamically active (i.e., objects do not have well-defined orbits and will eventually leave; \citealt{LevisonDones2007}). The Oort cloud likely formed as a result of migration of the giant planets, which caused surrounding small bodies to be scattered erratically and far out during early stages of arrangement of our planetary configuration \citep{Fernandez84,Nesvorny2018}. The scattered disk is thought to result from scattering events of Kuiper belt objects with Neptune and other giant planets (after the formation of the Kuiper Belt, e.g., \citealt{Pirani2021}). Long-period comets stem from the Oort cloud. Models suggest that $\sim3\%$ of cometary bodies in the Kuiper belt were originally trapped from the Oort cloud as well \citep{Nesvorny2017b}. Jupiter-family comets (JFCs) are short-period comets with periods of no more than $20$~years, whose current orbit is determined by Jupiter, but with origins in the Kuiper belt (not yet clear if dominated by the Scattered disk; \citealt{Duncan2004, VolkMalhotra2008}). This relative framing matters as different cometary origins might trace different physical radii within the natal solar protoplanetary disk, thereby tracing the conditions of that location at cometary birth. However, it remains difficult to pinpoint exactly which disk radii the different cometary family’s sample, beyond the fact that Oort cloud comets are more pristine (less thermally processed) than JFCs.

Cometary comae are typically water-dominated and boast a chemically rich volatile inventory. This is amply discussed in the reviews by \citet{MummaCharnley2011} and \citet{Altwegg2019}. The relative ratio of volatiles and refractories is a fundamental parameter to characterize. It is important to realize that the dust-to-water mass ratio does not equal the dust-to-volatiles mass ratio nor the refractory-to-ice mass ratio.  For the most well-studied comet, 67P/Churyumov-Gerasimenko, these three ratios span the $0.64-7.5$ range (\citealt{Choukroun2020} and the references therein), which makes it impossible to judge whether the comet is an ``icy mudball'' or a ``dirty snowball''. An accurate estimate can be considered the mission-integrated refractory-to-ice mass ratio of the nucleus obtained with the RSI instrument of $3-7$, which implies that the nucleus of 67P/C-G is a highly porous, very dusty body with very little ice \citep{Paetzold2019}.

Irrespective of the current dynamics of any particular comet, all comets are remnants of past planetesimal formation processes that took place in the protoplanetary disk that birthed our planetary system. An infant star-forming region does not contain any comet-sized objects prior to the onset of star and planet formation. The earliest phases contain dust grains predominantly smaller than $0.3~\mu$m \citep{Weingartner01}, which must assemble into objects with sizes of a few km (comet-size) to several $1~000$s of km (planet-size). Although the exact sequence of steps in this assembly remains to be elucidated, in one way or another comets must either be steppingstones or by-products of the physical processes that took place. Thanks to their current large distances from the Sun, and potentially early scattering into the outer sectors of our solar system, comets are the most pristine remnants of the protoplanetary disk \citep{AHearn2011b}. Consequently, placing stringent constraints on the physical properties and chemical compositions of comets are of utmost relevance for understanding the original buildings blocks of our solar system.

\subsection{Chondritic meteorites } 

{The chondritic meteorites are composed of three basic nebular components: chondrules, refractory inclusions (such as Ca-Al-rich inclusions or CAIs), and fine-grained matrix \citep{RN7147}. Chondrules and refractory inclusions are the products of relatively brief high-temperature ($\sim$1500-2100 K) processes. The fine-grained matrix is a complex mixture of materials, many of which were thermally processed in the solar nebula (e.g., crystalline silicates), as well as lesser amounts of materials like organic matter and presolar circumstellar grains that were not and were inherited from the protosolar molecular cloud. Based on their chemical, isotopic and physical properties, the chondrites have been divided into four classes (ordinary, carbonaceous, enstatite and Rumuruti), and subdivided into several groups (ordinary - H, L, LL; carbonaceous – CI, CM, CV, CO, CK, CR, CB, CH; enstatite – EH, EL). It is generally assumed that each group is derived from a separate parent body, but it cannot be ruled out that multiple parent bodies formed with similar compositions/properties.}

{The chondrites formed as unconsolidated `sediments' between $\sim$2 Ma and $\sim$4 Ma after CAIs (the oldest dated solar system objects). Various lithification processes (geologic mechanisms for generating a rock from a loose collection of grains) operating in their parent bodies after accretion produced `rocks' that were strong enough to survive impact excavation, and atmospheric entry. Lithification was driven largely by internal heating due to the decay of $^{26}$Al (t$_{1/2}\sim$0.7 Ma), although impacts may also have played a role. In asteroids that formed at $\sim$2 Ma (ordinary, Rumuruti, enstatite, CK), only meteorites that were excavated from close to their parent-body surfaces will have escaped severe heating and preserved some primitive materials. Except for the enstatite chondrites, all these meteorite parent bodies accreted water-ice, but in most meteorites this water has either reacted by oxidizing Fe-metal or been driven off by the internal heating. Parent bodies that formed between 3-4 Ma (e.g., CI, CM, CR) had sufficient $^{26}$Al to melt water-ice and drive reactions that generated hydrous minerals, carbonates, and Fe,Ni-oxides, but generally temperatures probably never exceeded $\sim$150°C. Planetesimals that formed after $\sim$4 Ma would not have had enough $^{26}$Al to even melt ice and are unlikely to be the sources of meteorites, unless heated by impacts, but could be sources of some interplanetary dust.}

{Dynamical simulations indicate that in the early solar system the asteroid belt was a region of stability that accumulated planetesimals that formed over a wide range of orbital distances and had been scattered there, mostly by the giant planets. Based on small, systematic isotopic variations in multiple elements \citep[e.g., Ca, Ti, Cr, Mo, Ru;][]{RN4486} recording the diversity of stellar nucleosynthetic processes (r-, s-, p- processes and so forth), the chondrites and other objects are now often classified as either carbonaceous or non-carbonaceous (the latter including Earth, Moon, and Mars). The association of the non-carbonaceous group with Earth and Mars, along with the need to keep the carbonaceous and non-carbonaceous groups separate, has led researchers to propose that the non-carbonaceous group formed inside of Jupiter’s orbit and the carbonaceous group beyond its orbit \citep{RN6896, RN7232, RN4486, RN7661}.}

{The accretion of ices, organics, and presolar materials by chondrites, and in the case of the carbonaceous group formation in the outer solar system, all point to genetic links between chondrites and comets. The ability to analyze large samples in the laboratory means that chondrites can provide important information about the nature of the dust in the solar nebula that cannot be inferred from remote observations of comets. However, the likelihood that lithification processes in even the most primitive chondrites have modified to some degree the primordial materials that they accreted must always be borne in mind.}

Two additional sources of extraterrestrial materials on Earth are interplanetary dust particles (IDPs) and micrometeorites. The definitions for these two types of particles are somewhat vague, but generally IDPs are taken to be <50-100 $\mu$m across, while micrometeorites are more in the 100-1000 $\mu$m range in diameter. IDPs with broadly chondritic compositions are divide into two categories, smooth (CS) and porous (CP) \citep{RN5739}. CS-IDPs are dominated by clay minerals, but also contain carbonates and other minerals that are characteristic of aqueously altered meteorites, although whether they are genetically related to known meteorite groups is uncertain. CP-IDPs are anhydrous, very fine grained and heterogeneous, and are generally assumed to have cometary origins.

Most IDPs were heated to at least 500~$^{\circ}$C for a few seconds during atmospheric entry. The larger micrometeorites were generally heated more severely during atmospheric entry, exhibiting a range of textures from mildly heated to fully molten and partially evaporated. Micrometeorites have many mineralogical, textural, and chemical properties that resemble those of the major chondrite groups, especially the carbonaceous chondrites \citep{RN8398}. However, the ultracarbonaceous Antarctic micrometeorites (UCAMMS) are quite distinct and are believed to be cometary \citep{RN4156}. As their name implies, they are dominated by carbonaceous material with variable D and $^{15}$N enrichments, but also tend to contain other primitive materials such as presolar grains \citep{RN4474,RN4156}. In any case, dynamical analyses of the zodiacal cloud dust suggests that a significant proportion of micrometeorites originate from comets \citep{Nesvorny2010}.

\subsection{Comets in the lab} 
 The Stardust mission brought back to Earth several micro-grams of dust collected in the coma of the JFC Wild 2 \citep{Brownlee2014}. This material is the only \textit{bona fide} cometary material that has been studied in the lab so far. To the surprise of the investigators, CAIs and chondrules that are mineralogically and chemically similar to those found in carbonaceous chondrites were both found in the Stardust samples \citep{Brownlee2014}. Furthermore, the O isotopic compositions of Wild 2 CAIs and chondrules are similar to those of their counterparts in carbonaceous chondrites \citep{Nakamura2008,McKeegan2006,Defouilloy2017}. In addition, the mineralogy of the Wild 2 fine-grained fraction has been found to bear similarities with carbonaceous chondrites' matrices and IDPs \citep{Zolensky2008, Ishii2008,Zhang2021}. Finally, cubanite (CuFe$_2$S$_3$) and magnetite (Fe$_3$O$_4$)
 have been found among Stardust samples \citep{Berger2011,Hicks2017}. Because these secondary minerals are typical of CI chondrites that have endured extensive aqueous alteration \citep{King2020}, the presence of these minerals among Stardust samples have been interpreted as hints of fluid circulation in the Wild 2 comet \citep{Berger2011,Hicks2017} (though magnetite can also be a high-temperature product). Given the absence of phyllosilicates, the intensity of aqueous alteration would have to have been far more limited than in CM, CR or CI chondrites \citep{Brownlee2014}. 
 
 Despite the violent capture process of the dust particles by the Stardust instrument, some organic particles were also found in the Stardust samples. The analysis of these particles proved to be very challenging, and it is not clear to what degree they were modified during their capture. Nevertheless, they were found to be heterogeneous, and compared to organics in chondrites and IDPs tended to be less aromatic, more O- and N-rich \citep{RN3171, RN3098}, and have smaller D and $^{15}$N enrichments \citep{McKeegan2006, RN4068}.
 
\begin{table*}[ht!]
\begin{center}
\caption{List of Probes of Birth Environment}
\includegraphics[width=15cm]{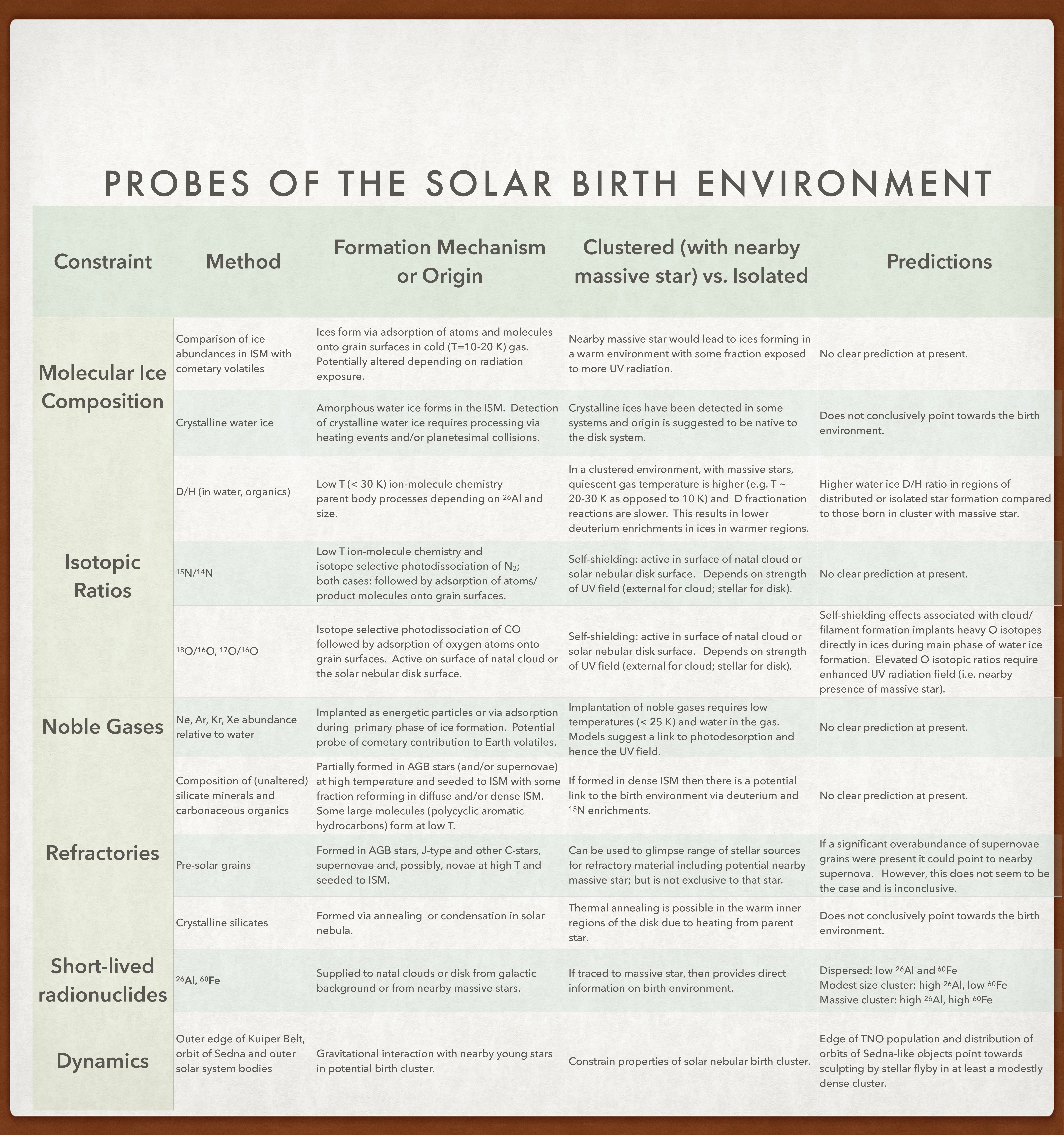}
\label{tab:obs}
\end{center}
\end{table*}
 
The fact that the cometary icy phase is so evidently visible to the public and astronomers alike, in the form of a spectacular coma, led to the concept that meteorites, which contain no ice and little water, could not come from comets. Meteorites have long been thought to originate from the ice-poor, inner solar system and therefore asteroids \citep{Kerridge88}.  Another obstacle for the acknowledgment of cometary meteorites is that comets have long been  perceived as primitive objects with components directly inherited from the ISM, while asteroids have been envisioned as processed solar system objects. This view has been challenged at the beginning of the 21$^{st}$ century by \citet{Gounelle2006} who established a cometary  orbit for the highly processed Orgueil CI1 meteorite. Building on that observation and other arguments, these authors proposed that rather than a rigorous dichotomy, there existed a continuum between low-albedo asteroids and comets \citep{Gounelle2008b}. The asteroid-comet continuum view is now gaining acceptance \citep{Hsieh2017, Engrand2016}, especially since small bodies such as comets and asteroids are known to have considerably moved between inner and outer solar system \citep{Levison2009} and because some asteroids are now considered to be volatile-rich \citep{Platz2016,Nuth2020}. If there is a continuum between dark asteroids and comets, there is possible that some carbonaceous chondrites might originate from cometary bodies as mentioned above. Most low-albedo asteroids might come from cometary regions \citep{Levison2009}. Indeed, the results from the Stardust cometary missions support a strong link between carbonaceous chondrites and comets \citep{Brownlee2014, Berger2011}. Moreover, it has been proposed that all the so-called carbonaceous meteorites \citep{RN4486, Kleine2020, RN7661} formed beyond the orbit of Jupiter.  This implies that at least some meteorites could come from zones of the nebula associated with ice-rich material (i.e. cometary formation zones). Perhaps the once hot debate about the possible existence of cometary meteorites has now abated, as it appears more and more evident that comets might not be as primitive as once thought. Also, getting closer to comets through the advent of space missions, it is realized that there is probably as much diversity among comets as among asteroids. So, it is very possible that not only the CI1 chondrites come from comets, but other meteorites groups originate from comets  \citep{VanKooten2011}.

\subsection{Summary of observational constraints on the birth environment}

Throughout this review we will explore the perspective offered by laboratory analysis and observational results of cometary refractories and volatiles.   In Table~\ref{tab:obs}, we summarize these aspects with a brief pointer towards the potential connection to the birth environment of the solar system within a Galactic context of star formation. We focus on several of these where substantive work has occurred and where connections to the birth environment can be made.  These are discussed in detail in Sections 2 and 3.  This is not fully inclusive of the list in Table~\ref{tab:obs}.  To be cover areas we do not discuss we provide some key references here.  For Nobel gases for instance the work of \citet{Monga15}, \citet{Marty17}, and  \citet{Ciesla18} deserves mention, while for crystalline ices we refer the reader to \citet{McClure15} and \citet{Min16}.








\begin{table}
\begin{center}
\caption{The principal nucleosynthetic reactions and sources of the isotopes of H, C, N, O, Mg, Si, and S \citep{Clayton2003}. Supernovae are abbreviated by SN and asymptotic giant branch stars by AGB.}
\label{tbl:minor_isotopes}
\begin{tabular}{rrr}
\hline
  & \textbf{Nucleosynthesis} & \textbf{Sources}\\
\hline
$^{1}$H & Big Bang & Big Bang\\
D & Big Bang & Big Bang\\
$^{12}$C & He-burning & Type II SN, AGB\\
$^{13}$C & CNO cycle & AGB\\
$^{14}$N & CNO cycle & AGB\\
$^{15}$N & CNO cycle & Novae, Type II SN, AGB\\
$^{16}$O & He-burning & Type II SN\\
$^{17}$O & CNO cycle & Novae, AGB\\
$^{18}$O & He-burning & Type II SN\\
$^{24}$Mg & C- and Ne-burning & Type II SN\\
$^{25}$Mg & He-burning & AGB\\
$^{26}$Mg & He-burning & AGB\\
$^{28}$Si & O-burning & Type II SN\\
$^{29}$Si & Ne-burning & Type II SN\\
$^{30}$Si & Ne-burning & Type II SN\\
$^{32}$S & O-burning & Type II SN\\
$^{33}$S & O-burning & Type II SN\\
$^{34}$S & O-burning & Type II SN\\
$^{36}$S & C-burning & Type II SN\\
\hline
\end{tabular}
\end{center}
\end{table}

\section{The galactic perspective of the origins of star and planetary systems}

\subsection{The Interstellar Medium}

The initial stages of cometary material begin in the interstellar medium (ISM) through the cycle of stellar birth and death. Stellar winds and supernova explosions return material to the ISM, which is then available for the formation of the next generation of stars and planetary systems in the Galaxy. We divide our discussion of the ISM into two components: dust and gas. In the following, we will distinguish between different components of the ISM based on density.

\subsubsection{Interstellar gas}
 
 The majority of the mass of the ISM resides in hydrogen with dust comprising only 1\% of the total mass.  The galactic ISM is comprised of several well-characterized components (hot ionized medium, warm neutral medium, warm ionized medium, cold neutral medium). The isotopes of most elements heavier than H and He are produced in the interiors of stars that are more massive than our Sun. The nucleosynthetic pathways for the elements that are most relevant to this chapter are summarized in Table~\ref{tbl:minor_isotopes}, together with the mechanisms that disperse them into space.  The overall gradient of past star formation rates in the Milky Way has created a dependence of heavy elemental abundance that declines with increasing galactocentric distance \citep{Wilson94}, but also must have local deviations.

There are known depletions of heavy elements
from the gas \citep{Savage96}, when assuming an abundance standard associated with the Sun or nearby B stars \citep{Asplund09, Nieva12}. This notably includes Fe, Si, Mg, and some O, and these elements are likely major constituents of interstellar dust.  About 50\% of elemental C and $<$18\% of N is missing from the gas \citep{Jensen07, Mishra15, Rice18}.

\begin{figure*}[ht!]
\begin{center}
\includegraphics[width=15cm]{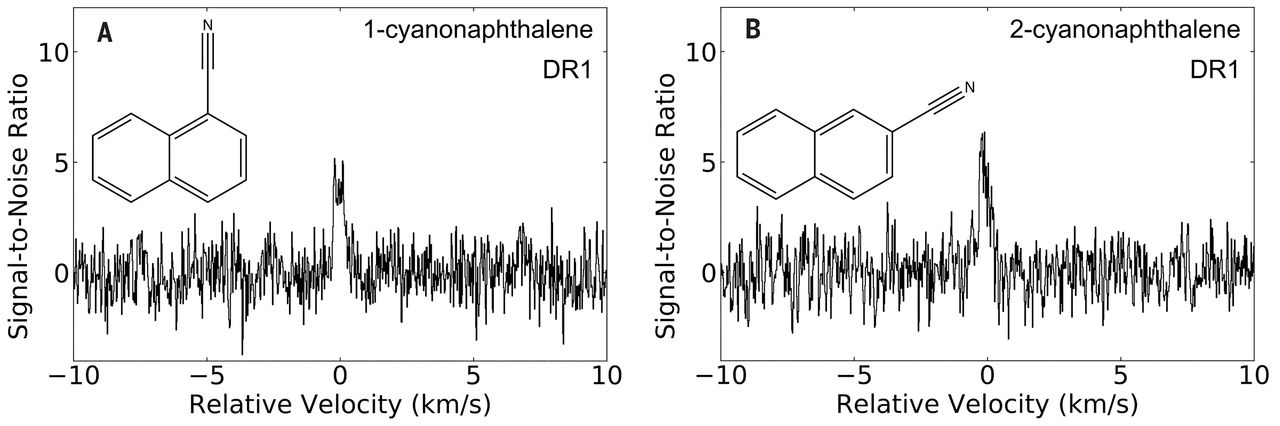}
\caption{Spectroscopic detection of 1- and 2-cyanonaphthalene towards the Taurus Molecular Cloud.  Taken from \citet{McGuire21}. }
\label{fig:pahs}
\end{center}
\end{figure*}

\subsubsection{Interstellar dust}

Late-stage stellar evolution is thought to play a key role in the formation of interstellar dust grains \citep{Henning10, Cherchneff13, Gobrecht16, Sarangi18}.
There is substantial evidence for the presence of dust grains in the galactic ISM based on the wavelength dependence of extinction \citep{Mathis90, Draine03} along with the aforementioned gas-phase depletion of heavy elements \citep{Savage96}.  Interstellar grain models generally assume a steep size distribution following size a$^{-3.5}$ \citep[labelled as MRN based on the last names of the three authors:][]{Mathis77}, and two compositions with contributions to extinction from silicates and carbonaceous materials \citep{Weingartner01}.   Models of carbonaceous grains have both aliphatic (carbon atoms in open chains) and aromatic (carbon atoms present in rings) components \citep{Jones13, Chiar13, Jones14}.  These are associated with broad spectral features seen in the diffuse ISM  \citep[aliphatic;][]{Gunay18} and, for aromatics, in gas in close proximity to massive stars or other sources of ultraviolet emission \citep[e.g., emission from polycyclic aromatic hydrocarbons, PAHs;][]{Tielens08}.

The external sources of ISM silicate dust are evolved star envelopes, winds, and ejecta with elemental C/O<1, where dust grains form by condensation from the hot ($>$ 1000~K) gas as it expands and cools.  This is supported by the fact that a fraction ($\sim$20\%) of silicates have crystalline structures and these are directly observed in circumstellar shells \citep{Molster02, Gielen08}. Carbonaceous grains form by analogous processes in stellar sources with elemental C/O $> 1$.

In the ISM there is processing of dust, as evidenced by the lack of crystalline silicates that is likely to be at least in part due to amorphitization of crystalline silicates in shocks or via interactions with energetic particles \citep{Demyk01, Kemper04}. Processing can also be much more destructive as ISM dust grains are subject to a variety of erosive processes that reduce sizes or even destroy them.  These include shocks \citep{Jones94, Micelotta10}, erosion in very hot (5000-6000~K) gas \citep{Bocchio12}, and cosmic-ray processing \citep{Micelotta11}. Estimates of the timescale of dust production in Asymptotic Giant Branch (AGB) stars and overall destruction in supernova shocks find a general mismatch in that rates of destruction exceed production, requiring an additional means of production \citep{Dwek98, Jones14}. Supernovae have been proposed as major contributors to dust production \citep{Dwek98, Sugerman06, Zhukovksa16}, which is supported by the detection of dusty galaxies at high ($z$ $>$ 6) redshift \citep{Bertoldi03, Watson15} when AGB stars do not have time to evolve.   Dust is clearly present in young supernova remnants, such as 1987A \citep[e.g.,][]{Matsuura15}; however, there are significant uncertainties in the yield  \citep[see these extensive reviews:][and references therein]{Sarangi18, Micelotta18}.  There remains considerable uncertainty and some dust production in the denser ISM could be required \citep[e.g.,][]{Rouille14}.  For example, \citet{Krasnokutski14} have shown experimentally that SiO clusters, etc., form at very low temperatures near absolute zero (it is barrierless), making grain formation in the ISM possible.


More recently, cm-wave observations have unambiguously detected emission from identified PAH molecules (1- and 2-cyanonaphthalene and others) forming {\em in situ} in dense ($n \sim 10^4$~cm$^{-3}$), cold ($T \sim 10$~K) dark molecular cloud cores \citep{McGuire18, Burkhardt21a, Burkhardt21, McGuire21}.  Concurrently, \citet{Cernicharo21} discovered o-Bezene and a host of other building block species \citep{Cernicharo21a}.  This exciting discovery is shown in Fig.~\ref{fig:pahs} and is enabled by a close coupling of laboratory spectroscopy and astronomical observations at cm wavelengths \citep{McCarthy21}.    The formation route for PAHs in the dense cloud environment is currently uncertain \citep{Burkhardt21}, but is unambiguously gas-phase in nature.    Thus, some fraction of large molecules/small grains are created in the dense ISM.   This is clearly important for cometary ices as these PAHs likely coat the surface of dust grains and are supplied to the planet-forming disk via collapse.   At present, it is not clear whether there is feedback of these cloud-created PAHs into the overall ISM as a dust production term.


\subsection{Molecular cloud, star, and planetesimal formation from the diffuse ISM}

\begin{figure*}[ht!]
\begin{center}
\includegraphics[width=15cm]{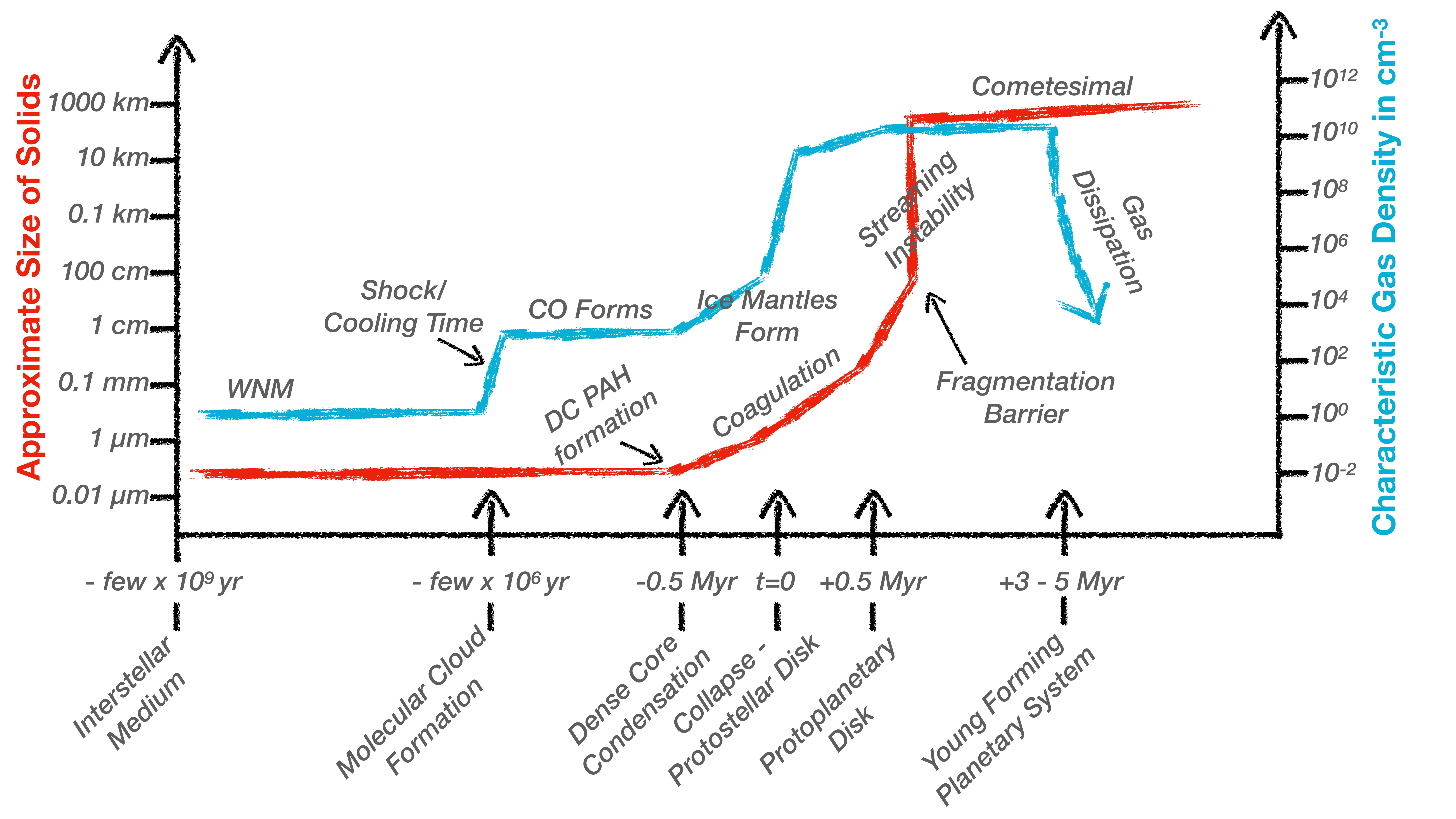}
\caption{Generalized schematic of solid growth (from $\mu$m to km) and characteristic gas density (in cm$^{-3}$) tracing the formation of cometary bodies in a natal disk. In this figure WNM refers to the Warm Neutral Medium phase of Galactic hydrogen and DC PAH formation relates to formation of polycyclic aromatic hydrocarbons in quiescent Dark Clouds. }
\label{fig:journey}
\end{center}
\end{figure*}

The journey of pre-cometary material, as shown schematically in Fig.~\ref{fig:journey}, starts in the low-density warm neutral medium (density or $n \sim 1$~cm$^{-3}$, temperature or T $\gtrsim$ 1000~K),  which comprises a large fraction of the hydrogen mass of the galactic interstellar medium \citep{Wolfire03, Heiles03}.   
ISM grains exist in this gas with an average size estimated via extinction of starlight to be of order 0.1 $\mu$m with $<\!n_{\rm gr} \, \sigma_{\rm gr}\!> \, \sim \, 2 \times 10^{-21} \, ( n / 1 \, {\rm cm}^{-3} ) \, {\rm cm}^{-1}$. Here, $n_{\rm gr}$ is the space density of grains, $\sigma_{\rm gr}$ is their cross-section. These grains remain bare due to the long collision timescales of atoms with grains ($t_{\rm gas-gr} = 1 / (n_{\rm gr} \, \sigma_{\rm gr} \, v) > 10^8 \, {\rm yr}$, where $v$ is the velocity of the gaseous atom or molecule) and the fact that abundant ultraviolet (UV) photons will desorb any atom that reaches the grain surface.   In this medium most of the H is atomic, with other elements similarly as atoms that are either neutral (e.g., O, N) or singly ionized (e.g., C, Fe) depending on the first ionization potential relative to the Lyman limit of 13.6 eV.   The galactic ISM is highly dynamic, and material becomes compressed behind colliding flows of gas or due to a shockwave induced by a nearby supernova \citep[see][and references therein]{Ballesteros-Paredes20}.  The shock-heated gas cools down on a cooling timescale (of order 1 Myr, but depends on shock strength) leading to a higher density and colder medium \citep{Clark12}.   Within this denser (n $>$ 100~cm$^{-3}$) medium H$_2$ forms via catalytic chemistry on dust surfaces and CO subsequently forms as the UV background becomes diluted, due to strong UV photo-absorption by dust gains and self-shielding by CO molecules, within the dense parcel of gas \citep{Bergin04, Glover10}.   The presence of CO denotes the observational presence of a molecular cloud. 

This molecular cloud is dense enough such that atoms are colliding with grains as the formation of the H$_2$-dominated cloud requires grain-surface formation of H$_2$.  Thus, a rudimentary grain mantle begins to form in this low density slightly warmer (20--30~K) gas \citep{Hassel10, Furuya15, Ruaud18}.  Simulations of this dynamic medium then show the formation of multiple filamentary systems with cores condensing within the filament (\S 3).  The increase in gas density associated within centrally concentrated dense core condensation leads to the rapid ($< 10^5$~yrs) formation of the ice mantles  that are widely characterized in interstellar space via observation at infrared wavelengths \citep[e.g.,][]{Boogert2015} and the creation of deuterium-enriched ices \citep{Ceccarelli14}.  This is the main phase of interstellar ice formation where key volatiles are formed and implanted (e.g. H$_2$O, CO, and CO$_2$).   Concurrently, there is evidence of modest grain growth in the centers of dense cores as suggested by \citet{Steinacker15} but see also \citet{Ysard16}.   

It is with the collapse of the core into the protostar and  protostellar disk that significant dust evolution occurs.  A central aspect is that the grains falling onto the disk from the surrounding envelope are covered in ices which appear to arrive  mostly unaltered \citep{Visser09, Cleeves14, Furuya16}.
In the disk, grains grow to mm sizes and settle to a dust-rich midplane quite rapidly \citep{Weidenschilling93, Pinte16}.  Observations, by \citet{vanthoff20}, are now demonstrating that protostellar disks appear to be warmer than the later stage protoplanetary disk (as the result of accretion), which might lead to some chemical alteration for hyper-volatile ices with the lowest sublimation temperatures \citep[e.g., CO and N$_2$;][]{Oberg21}.  Within the disk these mm-size pebbles are subject to differential forces that lead to inward drift \citep{Weidenschilling93, Birnstiel:2014}.  Thus, the outer regions of the natal disk may be an important source of material to the comet-forming zones. Ultimately, our current understanding is that the grains grow up to the fragmentation barrier, i.e., $\sim$cm-sizes \citep{Blum08, Birnstiel11}, and cometesimals form in regions of high dust density (low gas/dust ratio) where the streaming instability (or some other physical mechanism) is activated \citep{Youdin05, Johansen14}.

\subsection{Short-lived radionuclides}
\label{sec:SLR1}
The presence of radioactive elements in meteorites has long been known \citep{Urey,Lugaro:2018,Desch2022}. Some of them have short half-lives  compared to the age of the Sun and have now decayed \citep{2014Davis}. These so-called extinct short-lived radionuclides (SLRs) are identified through excesses of their decay products in meteorites \citep{1976Lee}. Long- and short-lived radionuclides provided  an internal heating source for early formed planetesimals. They can also be used for radiochronometric purposes, to probe the Galactic history of the Sun's building blocks and the astrophysical context of the Sun's birth. 

     The initial solar system abundances of radioactive elements are usually assumed to be that measured in CAIs, which are the oldest dated solar system solids \citep{Connelly12}. However, many SLRs have not been detected in CAIs and their initial solar system values are only inferred \citep{Dauphas:2011}. 
 Note that the very concept of $\it{an}$ initial value implicitly assumes the homogeneous spatial distribution of the considered SLR. Given that abundance variations can be either interpreted in term of spatial or temporal variations, it is difficult to firmly establish spatial homogeneity.  The existence of a significant portion of $^{26}$Al-poor CAIs among a majority of $^{26}$Al-rich CAIs seems to indicate that there was some degree of heterogeneity in the distribution of that important SLR \citep{2013Makide, Krot2019} whose half-life is 0.72 Myr. 
 
 
 Due to the difficulty of measurements and the scarcity of available material, no radioactive element has been detected in Wild 2 samples \citep{2010Matzel}. However, if (some) carbonaceous chondrites  come from comets, the presence of most radioactive elements in cometary regions is almost certain as most of the SLRs have been detected in carbonaceous chondrites.
 
 %


\subsection{What aspects of cometary material could trace the general ISM?}

\subsubsection{Initial ice mantles}

The molecular cloud phase ensues once the gas density increases to $> 10^{3}$~cm$^{-3}$ and the temperature drops to $<20$~K \citep[][]{BerginTafalla2007}. In this phase, simple molecules (H$_{2}$O, NH$_{3}$, CO$_{2}$, CH$_{4}$) start to be efficiently formed via grain-surface chemistry (e.g., \citealt{Hiraoka1998, Cuppen2017}) and the initial ice mantle forms (e.g., Fig.~\ref{fig:journey}). It has been experimentally demonstrated that under ISM-like physical conditions, H$_{2}$O is formed through hydrogenation of atomic O \citep{Dulieu2010}, NH$_{3}$ through the hydrogenation of atomic N \citep{Fedoseev2015a}, CO$_{2}$ through the association of CO and OH \citep{Ioppolo2011c}, and CH$_{4}$ through hydrogenation of atomic C \citep{Qasim2020}. Water is the dominant species forming the initial icy mantle on top of $0.1~\mu$m dust grains in molecular clouds owing to its high desorption temperature \citep{vanDishoeck2014a, vanDishoeck2021}. 

Deuterium formed alongside H in the Big Bang.    In the diffuse ISM, it is locked up in HD. Cosmic ray ionization of H$_{2}$ forms H$_{3}^{+}$ that in turn liberates D in the form of H$_{2}$D$^{+}$ for participation in chemical reactions.

\[ {\rm H_{2} \stackrel{C.R.}{\rightarrow} H_{2}^{+} \stackrel{H_{2}}{\rightarrow} H_{3}^{+} (+ H) \stackrel{HD}{\rightarrow} H_{2}D^{+} + (H_{2} + 230~K)}\] 
\[\stackrel{e^{-}}{\rightarrow} H_{2} + D\]

H$_{2}$D$^{+}$ paves the way for gas-phase molecules to become deuterated in particular at temperatures $<20$~K, as its production is strongly favored due to the $230$~K exothermicity of the reaction (the zero-point energy of the H-D bond is lower than that of H-H; \citealt{Watson1974}). Meanwhile, the final step of dissociative recombination with electrons increases the availability of atomic D in the gas. Upon adsorption onto grain surfaces, incorporation of D into solid-state molecules increases \citep{Roberts2003}. Deuterium readily participates in addition reactions on grain surfaces thanks to its longer residence time on the grains in comparison to that of the lighter, more mobile H \citep{Tielens1983}. Deuteration of molecules is particularly enhanced once CO freezes out on grain surfaces, impeding destructive gas-phase reactions with H$_{3}^{+}$ and H$_{2}$D$^{+}$ \citep{CaselliCeccarelli2012}. As the first H$_{2}$O ice layers form in molecular clouds, HDO is also synthesized via D addition reactions on grain surfaces.   These initial ice layers will have reduced D/H ratios compared to later stages associated with colder ($\sim10$~K) prestellar cores (\S 3.1).

Isotopic fractionation effects also occur for other major elemental pools: C, N, and O.   Of the major elements, H has the largest zero-point energy difference as the mass change between H and D is larger than for other isotopes (e.g., $^{12}$C vs. $^{13}$C, $^{16}$O vs. $^{18}$O, $^{14}$N vs. $^{15}$N).   Thus, the effects of fractionation via kinetic chemical reactions are reduced for the heavier elements.  Kinetic effects for C and O are discussed by \citet{Langer84}, \citet{Rollig13}, \citet{Loison19}, and \citet{Loison20}.
Nitrogen is an interesting case as clear evidence for fractionation exists in star-forming dense cores within nitriles (C-N bonds), nitrogen hydrides (N-H bonds), and potentially within N$_2$ \citep{Hily-Blant13, Redaelli18} along with $^{15}$N enrichments in meteorites \citep{Alexander12}.  However, the origin of these heavy isotope enrichments (and deficits) via gas-phase pathways in the dense ISM is uncertain \citep{Roueff15, Wirstrom18}.

 
Much larger fractionation effects occur in the gas phase, induced by UV photons and molecular self-shielding.   Most molecules have photo-dissociation cross-sections that are continuous across the UV spectrum \citep{Heays17}.  Because formation rates are less than photodissociation rates, grains must provide the shielding from UV.  H$_2$, CO, and N$_2$ are an exception in that their photodissociation cross-sections are through molecular bands or lines in discrete areas of the UV spectrum \citep{vandishoeck88}.\footnote{H$_2$O can self-shield in some instances due to fast formation rates \citep{Bethell09}.}  These spectral lines can be saturated with optical depths much greater than unity such that molecules downstream of the UV radiation could be self-shielded by the destructive effects of UV photons.   Isotopologues have slighted shifted spectral lines from the corresponding main species and reduced abundances; hence the onset of self-shielding occurs in somewhat deeper layers \citep{vandishoeck88, Heays14}.  For example, $^{12}$C$^{18}$O self-shields in deeper layers than $^{12}$C$^{16}$O.  This leads to a layer with excess $^{18}$O.  If this were to be incorporated into water ice via surface reactions it would carry an isotopic enrichment.  Self-shielding has been suggested as a mechanism to account O and N isotopic enrichments in meteorites \citep{Clayton93, Marty12}, which would be active on the surface of the molecular cloud to be provided by condensation first to the dense star-forming core and then by collapse to the young disk \citep{Yurimoto04}.  \citet{Lee:2008} explored the potential for cloud/core formation to contribute to O isotopic enrichments.  They find that UV radiation fields need to be elevated above the average interstellar radiation field for significant enrichments to be provided via collapse; i.e. it requires close proximity of a massive star.   But see discussion in \S 3.3 regarding protostellar disks as this is another potential area where self-shielding could be active \citep{Lyons05}.

\noindent {\it Summary of Potential Links to Birth Environment:}  The cloud could be the source of heavy isotopic enrichments due to the enhanced effects of radiation on lower density gas within molecular clouds (through self shielding).    This would imply the presence of a massive star and formation in a cluster (as opposed to isolation).  More work needs to be done to explore whether nitrogen might provide additional links and on the efficiency of similar processes with the young disk.


\subsubsection{Presolar circumstellar grains}
{Presolar circumstellar grains are found in the fine-grained matrices of the most primitive members of all the chondritic meteorite groups \citep{RN8795, RN6920, RN5821}, as well as in micrometeorites and IDPs, some of which probably come from comets. The other major components of chondrites, chondrules and refractory inclusions, formed at high enough temperatures in the solar nebula to have destroyed any presolar materials. Circumstellar grains are recognized as such based on their isotopic compositions, which are so different from solar system materials that they cannot be explained by typical physical and chemical processes that operated in the early solar system. Instead, their isotopic anomalies are best explained by nucleosynthetic processes that operate in evolved stars towards the end of their lives.}

{Based on their isotopic compositions, the dominant stellar sources are AGB stars, supernovae, J-type C stars, born-again AGB stars, and, possibly, novae. The number of stellar sources represented in the grains remains uncertain, but must be at least in the many tens \citep{RN644, RN1631}. However, the actual number may be much larger, as dating of the residence times of individual grains in the ISM suggests that they had been accumulating in the ISM for $\sim$1 Ga prior to formation of the solar system \citep{RN8213}.}

{The grains are typically micron to sub-micron in size, but grains up to 25 $\mu$m across have been found \citep{RN8547}. When found in situ in the meteorites, they are present as isolated grains that are mostly, but not always, composed of one phase. These grains almost certainly spent considerable time in the ISM but, contrary to the predictions of the Greenberg model for interstellar grains  \citep[e.g.,][]{RN3021}, they do not have observable carbonaceous layers surrounding them. The main presolar circumstellar grain phases are: silicates (both crystalline and amorphous), oxides (Al$_{2}$O$_{3}$, Mg(Al,Cr)$_{2}$O$_{4}$, Ca(Al,Ti)$_{12}$O$_{19}$, TiO$_{2}$), SiC, Si$_{3}$N$_{4}$, nanodiamonds, and graphite.}

{All the presolar grains are susceptible, to varying degrees, to destruction by the aqueous alteration and/or thermal metamorphism that lithified the chondrites. Nanodiamonds are arguably the most robust and, possibly, the most abundant type of presolar grain – their abundance is $\sim$1400 ppm in chondrite matrices. They are too small ($<$5 nm) to measure individually, but in bulk have Xe and Te isotopic compositions that suggest formation in supernovae. However, the abundances of Xe and Te are low and may only be carried by one in 10$^{5}$-10$^{6}$ grains. The bulk C and N isotopic compositions of the nanodiamonds are similar to those of the bulk solar values and, therefore, it is possible that the vast majority formed in the ISM and/or the early solar system. Nevertheless, to first order, the bulk abundances of the Xe-containing nanodiamonds are constant in the most primitive members of all chondrite groups, when corrected for their varying matrix contents.}

{The next most abundant and the most susceptible to destruction by parent body processes are the presolar circumstellar silicates. The highest silicate and oxide abundances ($\sim$175-250 ppm) have been found in the matrices of the most primitive CM, CO, and CR carbonaceous chondrites. The lower abundances reported for other members of these groups, as well as members of other chondrite groups, probably reflects the fact that they experienced more destructive parent body conditions. Presolar silicate and oxide abundances reported for CP-IDPs, the most likely IDPs to come from comets, range from 0 ppm to 15,000 ppm, with an average of 430 ppm \citep{RN7221}. This range must, to some extent, reflect the small areas analyzed for individual CP-IDPs, as well as variable atmospheric entry heating. However, it seems likely that CP-IDPs are composed of materials that experienced a range of reprocessing in the solar nebula and/or their parent bodies.}

{Based on their average presolar silicate and oxide abundances, IDPs contain roughly a factor of two more unprocessed molecular cloud material than chondrite matrices. Presolar silicate/oxide abundances have been measured in some unambiguously cometary material, i.e., returned Wild 2 material. This material was severely affected by the high velocity ($\sim$6 km/s) capture process. However, by comparing the survival efficiency of presolar grains in experiments designed to simulate the capture process with what is found in the returned samples, \cite{RN8792} estimated presolar circumstellar silicate and oxide abundances in Wild 2 dust of 600-830 ppm (i.e., $\sim$2.5-5 times what is found in primitive chondrite matrices). One other thing to bear in mind when discussing presolar circumstellar silicate/oxide abundances is that \cite{RN7310} argue that typical in situ searches for silicate/oxide presolar grains underestimate their abundances by at least a factor of two because detection efficiencies decrease rapidly forf grains with diameters below $\sim$225 $\mu$m.}

{Roughly 40$\%$ of presolar silicates found in meteorites and IDPs are crystalline \citep{RN7221}. This compares to the $<$2 $\%$ crystalline fraction in the diffuse ISM silicate dust \citep{RN3032, Kemper04}. The higher abundances of crystalline silicates observed in meteorites, IDPs and comets (e.g., Wild 2) than in diffuse ISM dust could have been produced by at least three processes: annealing, melting/crystallization, and condensation from a gas. Presumably this is also the case for extrasolar disks with crystalline silicates. All three processes for making crystalline silicates are likely to result in the O isotopic exchange of circumstellar grains with H$_{2}$O in the disk gas \citep[e.g.,][]{RN7711}, and in many circumstances with the more abundant interstellar silicate dust, which would have destroyed any nucleosynthetic isotopic signatures. Hence, the level of crystallinity of the presolar circumstellar silicates is almost certainly primary and is of roughly the same order as the 10-20$\%$ crystallinity estimated for silicates in stellar outflows \citep{Kemper04}. From the $<$~2$\%$ crystallinity of ISM dust and the 40$\%$ crystallinity of circumstellar presolar grains, we can place an upper limit of $<$~5.5$\%$ for unprocessed circumstellar silicates in the diffuse ISM, which would increase to $<$~8-18$\%$ if we use the crystallinity estimates for stellar outflows.

Other estimates suggest that the fractions of circumstellar grains in the diffuse ISM may be $\le$0.4-2 $\%$ \citep{RN7221} and a few percent \citep{RN7310}. These values are similar to some model predictions for grain destruction and reformation in the ISM that suggest that circumstellar grains make up $\sim$3 $\%$ of ISM dust \citep{RN8793}.}

{Assuming circumstellar grain abundances of 2-3$\%$ in the ISM means that 30-50 times as much interstellar dust must have accompanied the presolar silicates found in extraterrestrial materials. It also suggests that the $\le$250 ppm of presolar silicates in chondrite matrices represent only $\le$0.8-1.3 $\%$ of the original ISM dust from which the chondrite matrices ultimately formed, 1.4-2.2$\%$ for average CP-IDPs (430 ppm) and 2-4$\%$ for Wild 2 dust. These numbers would at least double if the \cite{RN7310} detection efficiencies are applicable to most in situ presolar silicate/oxide searches. Nevertheless, the numbers give some indication of the level of reprocessing of the original protosolar molecular cloud material these primitive materials experienced. Although interstellar silicate grains should vastly outnumber the circumstellar ones, unambiguously identifying them is problematic because, if they formed in the ISM, they are expected to have similar isotopic compositions to the bulk solar values.}

\noindent {\it Summary of Potential Links to Birth Environment:}  The circumstellar grains in primitive extraterrestrial materials provide constraints on the range and number of stellar sources that contributed material to the protosolar molecular cloud in the last billion years or so. Most of the grains formed around AGB stars and other relatively low mass, long-lived stars that are unlikely to have ended their lives in the protosolar molecular cloud. On the other hand, if a massive star, within a clustered star-forming environment, had ended its life as a type II supernova in the vicinity of the protosolar molecular, one might expect an excess in supernova grains in the circumstellar grain population. At present, unlike with AGB grains, it is not possible to estimate how many supernovae are represented in the circumstellar grain population.  In addition, estimates of supernova dust production rates are quite uncertain. Nevertheless, to date there is little indication that the solar system has a higher proportion of supernova-derived grains than expected for the solar neighborhood.

\subsubsection{Refractory organics in chondrites and comets}
{There are roughly equal masses of silicate and carbonaceous dust in the diffuse ISM \citep{RN5617, RN6212, RN2660}. Using the abundances of circumstellar silicates in ISM dust estimated above, this means that there should be roughly 30-50 times as much interstellar carbonaceous dust in the extraterrestrial materials as circumstellar silicates, or 0.75-1.3 wt.$\%$ carbonaceous dust in chondrite matrices, 1.3-2.2 wt.$\%$ in average CP-IDPs, and 1.8-4.2 wt.$\%$ in Wild 2 dust. Again, these abundances may at least double if the \cite{RN7310} circumstellar silicate detection efficiency estimates are typical for in situ searches. Primitive chondrite matrices contain $\sim$3-4 wt.$\%$ organic C, which is similar to the estimates for interstellar carbonaceous dust that should have accompanied the circumstellar silicates. This raises the question of whether some or all of the organic C in chondrites, and presumably comets, could ultimately have an interstellar origin. The C contents of CP-IDPs tend to be significantly higher than would be predicted from their average circumstellar grain abundances \citep{RN2365, RN1356}, but there is a strong bias in favor of low densities (e.g., organic-rich) in particles that survive atmospheric entry relatively unheated. The abundance of carbonaceous material in returned Wild 2 dust could not be determined because much of it is likely to have been destroyed by the capture process.}

{While primitive chondrites  contain complex suites of solvent extractable compounds, such as amino and carboxylic acids \citep{RN7566}, the bulk of the organic C in chondrites is insoluble in typical solvents and so is thought to be macromolecular \citep{RN7117, RN7566}. However, on average only about half of this C can typically be isolated from primitive meteorites for reasons that are not understood. The isolatable material is normally referred to as insoluble organic matter or IOM, and its properties and possible origins has recently been extensively reviewed by \cite{RN7117} and are summarized below.}

{In the most primitive chondrites, IOM has an elemental composition of C$_{100}$H$_{75-79}$O$_{11-17}$N$_{3-4}$S$_{1-3}$ (relative to 100 Cs). Studies in situ and of IOM isolates show that the organic C is present in grains with a wide range of sizes and morphologies, and that there is no obvious spatial relationship between the organics and any inorganic materials. Most organic grains are submicron in size, with larger particles that are found in situ being aggregates of smaller grains that may result from fluid flow concentrating grains in veins. The most distinctive grain morphologies, although not the most abundant, are those of so-called nanoglobules that are typically roughly spherical and hollow. Nanoglobule-like objects have also been found in IDPs and Wild 2 samples.}

{Infrared (IR) and nuclear magnetic resonance (NMR) spectroscopy indicates that the IOM is predominantly composed of small, highly substituted polyaromatic units that are linked together by short, highly branched aliphatic material. IOM is also very rich in D and $^{15}$N, with bulk compositions that range up to D/H$\approx$6.2-7.0$\times$10$^{-4}$ and $^{14}$N/$^{15}$N $\approx$190-240, and micron to sub-micron hotspots, probably associated with individual grains, that can range up to D/H$\approx$6.4$\times$10$^{-3}$ and $^{14}$N/$^{15}$N$\approx$70. There is no straightforward correlation between the extent of D and $^{15}$N enrichments, either in bulk IOM or in hotspots. The C, O and S isotopic compositions of IOM do not seem to be very different from terrestrial values. The D and $^{15}$N enrichments are generally interpreted as being due to the formation of the IOM or its precursors in cold and/or radiation-rich environments, although whether this was in the presolar molecular cloud, or the outer solar system remains the subject of debate.}

Comets 1P/Halley and 67P/Churyumov-Gerasimenko contain refractory (macromolecular?) carbonaceous material that is roughly as abundant by mass as the silicate dust \citep{RN7354, RN3820}. In 1P/Halley, the carbonaceous material has a bulk composition of roughly C$_{100}$H$_{80}$O$_{20}$N$_{4}$S$_{2}$ \citep{RN2077}, while in 67P it has a composition of roughly C$_{100}$H$_{104}$N$_{3.5}$ \citep[][the O and S contents were not measured]{RN7441, RN7889}. The refractory carbonaceous material in 67P is also very D-rich with a D/H=1.57$\pm$0.54$\times$10$^{-3}$ \citep{RN8727}. 
In other comets, the carbonaceous dust contents appear range from below those of 1P/Halley and 67P \citep{RN3366, RN3368} to comparable to them \citep{Woodward21}, perhaps reflecting a range of primitiveness. 


{The similar elemental compositions of IOM and refractory carbonaceous dust in comets 1P/Halley and 67P suggest that there is a genetic link between these materials. The higher abundance of the carbonaceous material in the two comets, along with the higher H/C and D/H, all point to the dust in these comets having been less processed in the solar nebula or in their parent bodies. Indeed, the abundances of carbonaceous material in 1P/Halley and 67P is roughly consistent with their dust being predominantly composed of unprocessed molecular cloud material. If the carbonaceous material is not interstellar in origin, then mechanisms must be found that destroy the interstellar carbonaceous dust and then remake carbonaceous dust, at least in the outer solar system, very efficiently.}

{If there is a genetic relationship between IOM and the carbonaceous material in comets 1P/Halley and 67P, a possible interstellar connection can be explored by comparing the properties of IOM with what is known of carbonaceous dust in the ISM. Some of the early speculation that IOM might be related to interstellar carbonaceous dust came with the recognition that there is a striking resemblance between the 3-4 $\mu$m IR spectra (the aliphatic C-H stretch region) of IOM and diffuse ISM dust \citep{RN3745, RN1633, RN1628}. \cite{RN2363} interpreted the 3-4 $\mu$m IR spectra of diffuse ISM dust as being due to short aliphatic chains attached to electronegative groups, such as aromatic rings, O and N. Subsequently, \cite{RN2495} concluded that the diffuse ISM dust contains few, if any, heteroatoms (e.g., O and N) and that the aromatic units must be large. They also pointed out that a range of materials give reasonably good fits to the diffuse ISM spectra, not just IOM.}


{The short aliphatic chains attached to aromatic units is certainly reminiscent of the structure of IOM, but a predominance of large polyaromatic units is inconsistent with what is seen in IOM. On the other hand, other models for the structure of diffuse ISM carbonaceous dust prefer small aromatic units \citep{RN2662, RN3765, RN4464}. The absence of heteroatoms in diffuse ISM carbonaceous dust is, potentially, more problematic given the abundance of O, in particular, in IOM and 1P/Halley carbonaceous dust. Isotopic compositions also pose a challenge to there being a link between diffuse ISM dust and IOM, as well as the refractory carbonaceous dust in 67P. Carbon stars are thought to be the major stellar sources of carbonaceous dust in the diffuse ISM. As a result of nucleosynthesis and dredge-up in these stars, their dust should be depleted in D and $^{15}$N and exhibit a very wide range in C isotopic compositions (as is seen in presolar circumstellar SiC and graphite grains). This is the opposite of what is seen in IOM (and 67P dust), which is D and $^{15}$N enriched and has normal C isotopic compositions. The C isotopes of IOM would be consistent with efficient destruction of circumstellar dust and reformation in the ISM, but the enrichments in D and $^{15}$N are difficult to explain via isotopic fractionation in the diffuse ISM.}

{However, the solar system formed from molecular cloud material not diffuse ISM material. Is it possible that diffuse ISM dust is modified in molecular clouds to look more like IOM and the refractory carbonaceous material in comets 1P/Halley and 67P? The detection of PAHs forming in cold conditions (\S 2.2) where D-enrichments would be active is suggestive of a potential contributor. Another possibility is that radiation damage by cosmic rays accumulates in very cold, ice-coated carbonaceous dust grains while resident in molecular clouds \citep{RN3594}. These ices are likely to be enriched in both D and $^{15}$N. Upon warming during or after formation of the solar system, radiation generated radicals in the ice will become mobile and able to react with radiation damaged regions of the carbonaceous dust, thereby adding H, N, and O with their associated isotopic enrichments. Also, adding H to radiation damaged interior regions of large aromatic units would generate both smaller aromatic units and short, branched aliphatic chains. This mechanism for providing a direct path from diffuse ISM dust to the carbonaceous material in comets and meteorites remains speculative at present, but at least in principle could be experimentally tested.}

Other proposed mechanisms for making IOM are envisaged to have operated in the early solar system include Fischer-Tropsch-type  synthesis, but the inefficiency of FTT synthesis under solar nebula conditions is problematic.
Irradiation of ices is another potential source of condensed organic material \citep{RN4686}, but experiments on interstellar/disk ice analogs do not produce materials that resemble IOM or the refractory C in comets \citep{RN6275}. An alternative energy source would be irradiation by energetic particles and/or UV of ice-free carbonaceous particles \citep{RN4686}.

\noindent {\it Summary of Potential Links to Birth Environment:} While a solar system origin for the IOM cannot be ruled out, an interstellar origin seems more likely given the high abundance of IOM-like material in comets with volatile ice compositions that suggest a molecular cloud heritage. If the IOM did ultimately form in the protosolar molecular cloud, can it provide any constraints on the solar system's birth environment?  The detection of PAHs forming in dark clouds offers a potential avenue for molecular cloud origin of the IOM.  Although, the full import of this dark cloud PAH formation is currently uncertain, some aspects would be clear.  PAHs are forming in dark clouds at cold ($\sim$10--20~K) temperatures \citep{McGuire21, Cernicharo21} where deuterium fractionation reactions are active.  Thus, they will form with D-enrichments.  This formation will occur via reactions linking to CH$_2$D$^+$ \citep{Millar89}.  This pathway fractionates at higher temperatures than the reactions that form water, which are linked to H$_2$D$^+$ \citep{Roueff15}.  At face value this would imply efficient deuterium fractionation routes for organics with less temperature dependence (i.e., harder to link to subtle changes in the gas physical state of the birth environment).  Moreover, in the dark cloud environment nitrogen self-shielding may be active at modest extinctions \citep[$\sim 1.5^m$;][]{Heays14}.  Nitrogen-bearing PAHs are detected, and nitrogen is playing a part in the complex organic gas phase chemistry  \citep{Burkhardt21}.  Thus, $^{15}$N enrichments could be implanted as well. This would depend on the strength of the external UV field and, like oxygen \citep{Yurimoto04, Lee:2008}, the $^{15}$N enrichment could relate to the presence of a nearby massive star.  More work is needed to understand any potential predictions that might discriminate between various scenarios for the solar birth environment.

\section{The Local Birth Site}

\subsection{Star cluster dynamical perspectives: solar birth in a cluster?}
\label{sec:dynamics}
Most stars form as part of a stellar group \citep{Lada:2003, Porras:2003}, possibly this holds also for the Sun. The following features have been considered as  indicators of the Sun's cluster origin.

First, the there is a steep drop in surface density of the disc at \mbox{$\approx$ 30 au}. Beyond,  only the equivalent of 0.06 times the Earth's mass is contained in all Kuiper belt objects together \citep{Di_Rusco:2020}. It is likely that the solar system's disk was initially considerably more extended and then truncated \citep{Morbi:2004}.  Disk truncation could have happened by a close flyby of another star \citep{Ida:2000,Kenyon:2004,Pfalzner:2018,Batygin:2020}, external photo-evaporation by nearby massive stars \citep{Adams04, Owen:2010, Mitchell:2011,Winter18,Concha19,Concha21}, an earlier binary companion to the Sun \citep{Matese:2005,Siraj:2020} and a nearby supernova explosion \citep{Chevalier:2000}. All these processes require a strong interaction with neighboring stars, which is most easily realized in a cluster environment. 

Second, most Trans-Neptunian objects (TNOs) move on highly inclined orbits, e.g., Sedna 
\citep{Brown:2004,Trujillo:2014, Becker:2018}.  
Neptune could not have catapulted the Sedna-like objects to such distances, but an additional external dynamical mechanism is required. The close flyby of another star is one possibility. Sedna-like objects could either have been part of the primordial disk and flung out  \citep[e.g.,][]{Ida:2000,Kenyon:2004,Batygin:2020} or they were captured from the disk of the perturber star \citep{Jilkova:2015}. The periastron distances  would have been 50 -- 350 au  \citep{Jilkova:2015,Pfalzner:2018, Moore:2020}. Nowadays, such close flybys are infrequent for the solar system \citep{Bailer:2018}, but were more common earlier on if the Sun formed within a young cluster.  


Each of these features has been used to constrain the properties of the solar system birth cluster
\citep{Mitchell:2011,Pfalzner:2013,Li:2015,Jilkova:2015, Owen:2010, Pfalzner:2020,Moore:2020}. 
The general consensus is that clusters like Westerlund 2 or Trumpler 14 are excluded as solar birth environments due to their extremely high stellar density ($n >$ 10$^5$ stars pc$^{-3}$) and very strong radiation fields ($G \gg$ 1000) \citep[for example,][]{Hester:2004, Williams:2007, Adams:2010}. 
Dynamical and radiation arguments alike \citep{Adams:2010,Li:2015,Pfalzner:2020}, agree that it is unlikely that the solar birth cluster contained $N>$10,000 stars, because in such environments discs mostly are truncated to sizes well below 30 au. Similarly, there seems to be fair agreement on the lower limit independent of the constraining feature. 
Also, one cannot completely exclude that the Sun formed in a low-mass clusters \citep{Parker16}, an environment similar to the one recently suggested by \cite{Portegies:2019} with \mbox{$N$ = 2500 $\pm$ 300} stars and a radius of \mbox{$r_{vir}$ = 0.75 $\pm$ 0.25 pc} seems to be a more likely choice.

Recent studies find it more useful to define a typical local stellar density rather than limiting the number of stars. The existence of long- and short-lived clusters means that there exists no general one-to-one relation between the number of stars and stellar density \citep{Pfalzner:2009}. Equally, sub-structure in the cluster means that local densities can exceed average stellar densities considerably \citep{Parker:2014}. 
Simulations show that in 90\% of all cases solar system analogues are formed in areas where the local stellar density exceeds $\rho_{local} > $ 5 $\times$ 10$^4$ pc$^{-3}$ \citep{Brasser:2006,Schwamb:2010,Pfalzner:2020,Moore:2020}. Clusters with an average cluster density of 100 pc$^{-3}$ to a few 10$^4$ pc$^{-3}$ usually contain areas within this density range. In recent years, the first solar sibling candidates have been identified \citep{Ramirez:2014, Bobylev:2014,Liu:2016,Martinez:2016,Webb:2020}, strengthening the argument that the Sun was born as part of a cluster of stars.

\begin{figure*}[ht!]
\begin{center}
\includegraphics[width=15cm]{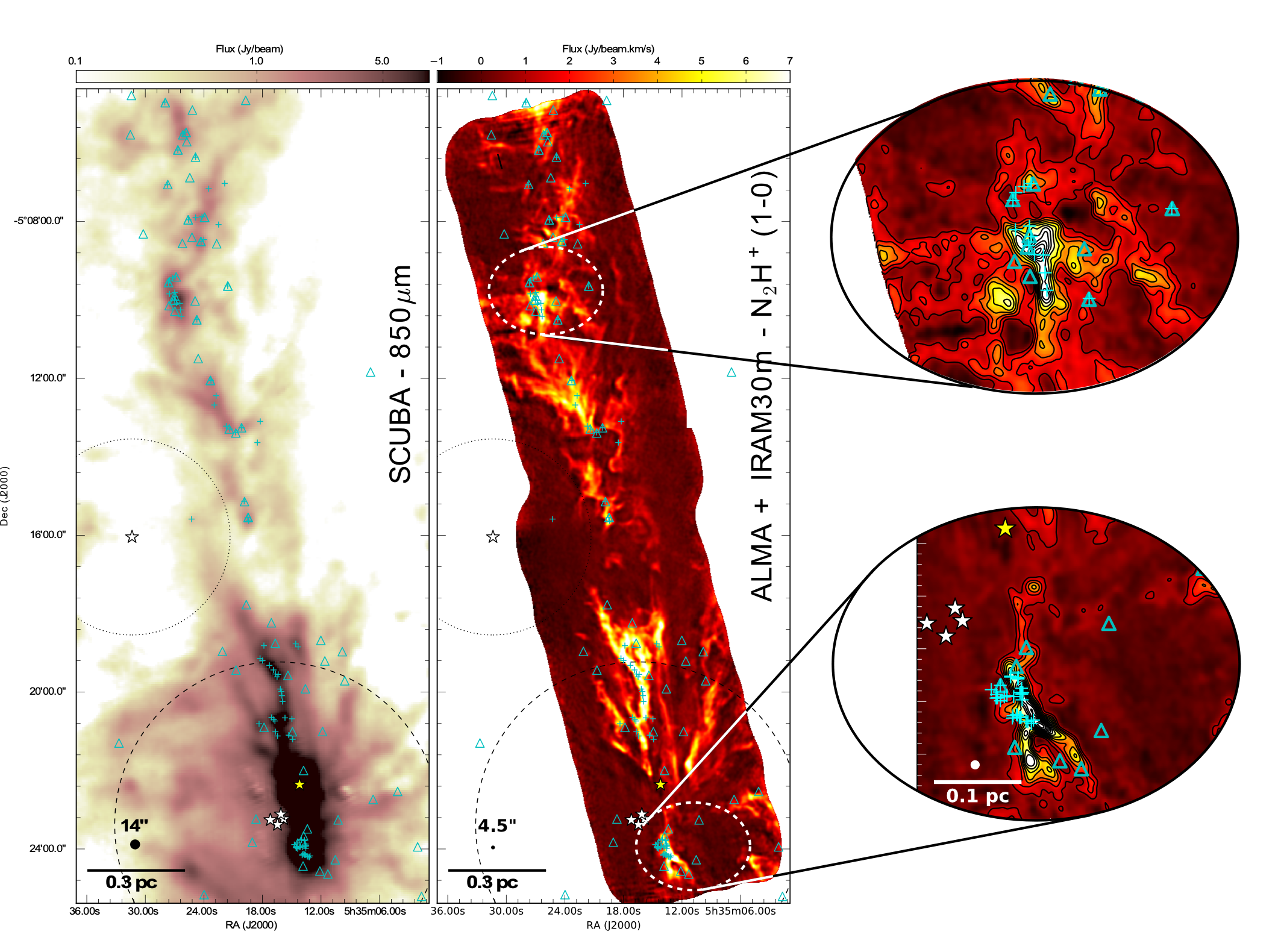}
\caption{N$_2$H$^+$ emission map of the Orion integral filament taken from \citet{Hacar18} which includes 850$\mu$m dust continuum emission from \citet{Johnstone99}.   In this figure the white stars are the positions of the Trapezium with the dashed line representing the 0.5 pc radius of the Orion Nebular Cluster, the open black star with circle shows the M43 nebula, the yellow star relates to the Orion BN source, and Spitzer protostars are shown as blue triangles.  The beam sizes of the images are given on the bottom right.}
\label{fig:orion}
\end{center}
\end{figure*}
\subsection{Dense filaments in molecular clouds}

\subsubsection{General structure and properties}

Wide field mapping of thermal continuum emission from dust by instruments such SCUBA and Herschel SPIRE/PACS show molecular clouds are dominated by dense webs of filamentary substructure \citep{Motte98, Andre10, Arzoumanian11, Andre14, Motte18}.  \citet{Hacar13} showed that filaments contain additional substructure (called fibers), in velocity, when observed using the best tracer of dense molecular gas, N$_2$H$^+$ \citep[for a  discussion of N$_2$H$^+$ see][]{BerginTafalla2007}.  A sample of this structure is shown in Fig.~\ref{fig:orion} where the thermal dust continuum image of the central regions of the Orion molecular cloud, the nearest site of massive star-cluster formation, is shown alongside an ALMA image of N$_2$H$^{+}$ emission with a spatial resolution of ~1800 au. For a larger-scale view of the filamentary structure in Orion, the reader is referred to \citet{Bally87}.  The images in Fig.~\ref{fig:orion} illustrate the general sense of star formation as a series of nested filaments, associated with active star-birth, converging towards hubs where clusters are born \citep{Myers09}. Another key perspective is that dense core collapse within filamentary structure could potentially lead to inhomogeneity within infall streams and different portions of the filament may contribute at different times to the disk \citep{Pineda20}.

In the denser (n $>$ 10$^4$ cm$^{-3}$) filaments, e.g., Fig.~\ref{fig:journey}, hundreds of ice layers are placed on top of the initial mantle formed with the cloud.  As the thickness of ice increases, the deuteration fraction of H$_{2}$O, and other carriers, will increase with dropping temperatures and a larger availability of atomic D. Further, the  freeze out of CO enhances the abundance of H$_2$D$^+$ leading this gas to be the main formation phase of D-enrichments in pre-cometary ices \citep{Taquet2014}.
The temperature of this gas matters.  Even a slight 5--10~K difference in temperature has a chemical effect on the composition of volatile ices.  For example, the D/H ratio of water ice would change by a factor of $\sim$3 between 15~K to 20~K \citep[see, e.g.,][]{Lee15}.

A central distinction between clustered star-forming environments compared and those that are isolated is that pre-stellar cores in Orion appear warmer (e.g., $> 10$~K) due to the presence of (passive) external heating sources \citep[e.g., nearby massive stars;][]{Harju93, Li03, Kirk17} and perhaps embedded in  filaments with higher average density \citep{Hacar18}. Herschel surveys of dust emission of nearby Gould Belt star-forming clouds \citep{Andre10} place this conclusion on sound statistical footing.  For example, material in Ophiuchus has a higher overall dust temperature \citep{Ladjelate20}, than Taurus \citep{Palmeirim13}.\footnote{The Herschel studies are generally average temperatures along the line of sight.  There does exist line of sight temperature structure such that colder gas exists in the deep interior of centrally concentrated cores \citep{Crapsi07, Roy14}.}
Clustered star-forming filaments also must have surfaces exposed to higher radiation fields as, for example, the average UV field along the Orion filament exceeds 1000$\times$ the average interstellar radiation field \citep{Goicoechea15, Goicoechea19}.  This is in stark contrast towards more isolated regions \citep{Goldsmith10}, where the UV field is closer to the average interstellar radiation field \citep{Habing68}.   This difference, and the effects of self-shielding, would alter the O and N isotopic ratios on cloud or disk surfaces, and potentially implant heavy isotopes into forming ice mantles \citep{Yurimoto04, Lyons05, Lee:2008}.

\subsubsection{Sources of short- lived radionuclides to molecular clouds}
\label{sec:SLRS2}

 The longer-lived radioisotopes in the early solar system (T$ >$  2.5 Myr) result from the continuous production (and decay) in the Galaxy \citep{Meyer2000} and will not be discussed here. This also the case for $^{60}$Fe (T = 2.62 Myr) whose abundance in the early Solar System is lower than initially thought \citep{Trappitsch2018}. Beryllium-10 (T = 1.39 Myr) has been suggested to have been inherited from the molecular cloud \citep{Desch2004} or made by irradiation in the protoplanetary disk \citep{McKeegan2000} together with $^{41}$Ca (T = 0.1 Myr) \citep{Liu2017} and $^{36}$Cl (T = 0.3 Myr) \citep{Tang2017}. The observed heterogeneity of $^{10}$Be  is incompatible with an inherited origin \citep{Dunham2022,Fukoda2019, Fukuda2021}.
 Aluminium-26, by far the most documented SLR,  is not made by irradiation \citep{Fitoussi2009}. Given its short-half-life compared to typical timescales of star formation, and its high abundance compared to the expectations of Galactic nucleosynthesis \citep{Meyer2000}, it likely requires a local last-minute stellar origin. If it had been inherited from the Galaxy as suggested by \citet{Young2016} and \citet{Gaidos2009}, it would not exhibit any spatial heterogeneity \citep{Krot2019, 2013Makide}. AGB stars and supernovae have been considered as potential candidates. Supernovae can now be excluded, because they would yield a $^{60}$Fe/$^{26}$Al ratio at least one order of magnitude larger to the one observed in the early solar system \citep{Gounelle2008}.
 
 The probability of associating an AGB star with a star-forming region is very small \citep{Kastner1994}. At present the most promising models are the ones involving massive star winds \citep{Arnould1997, Gounelle2012}. Iron-60 is indeed absent from winds, avoiding the supernova caveat. Because massive stars have short existence times and end their lives as supernovae, models that require a multiplicity of massive stars \citep{Gaidos2009} would also yield excesses of $^{60}$Fe relative to $^{26}$Al and what is observed in meteorites. 
 A setting whereby a massive star injects $^{26}$Al in a dense shell it has itself formed and within which a new star generation \citep{Gounelle2012, Dwarkadas2017} is formed is more likely than a runaway Wolf-Rayet star \citep{Tatischeff2010}. In such a context, the massive star source of $^{26}$Al would belong to a cluster of a few thousand stars, in rough agreement with dynamical constraints (see Section \ref{sec:dynamics}). 
Such a setting, though compatible with our present knowledge of star formation would concern only a few percent of stars \citep{Gounelle:2015} and has not been directly observed at present \citep{Dale2015}

\subsection{Protostellar disks and envelopes}

 Protostellar disks are a central phase of evolution as the early disk must have been hot enough to lead to the formation of CAIs and other refractory inclusions.  CAIs are the oldest dated Solar System materials \citep{Connelly12} and, assuming that the Al-Mg system can be used as a chronometer, formed over a period of order 10$^{5}$ yrs \citep{RN8990, RN3042}.   This and the high ($>$1400~K) formation temperatures suggest that the high temperature phase in the disk must be short lived or associated with accretion bursts (e.g., FU Ori events) \citep{RN4798, RN8250, RN8979, Kristensen18}.  Current data do show that protostellar disks are warmer than their protoplanetary counterparts, but the temperatures are still moderate \citep[tens to $\sim$100~K;][]{vanthoff20}.   This does not imply that hotter material does not exist. Rather this material is difficult to isolate via astronomical observation due to the presence of high dust optical depth, both in the disk and the surrounding envelope, and the existence of hot shocked gas in close proximity to the forming star. A hot phase associated with solid state (ice or refractory) sublimation must be important and would operate to reset the chemistry.  For example, the D-enrichments from ices would be erased \citep{Drouart99, Yang13} and  C-rich grains  irreversibly destroyed \citep{Li21}.  However, chemical reset cannot be absolute as infall, and the inwards drift of pebbles would supply fresh primordial material during subsequent evolutionary stages.  Much of the material exposed to the highest temperatures within a heating event will be accreted onto the star.

 Observations have gradually revealed kinematically distinct protostellar disks within embedded young protostellar envelopes \citep{Tobin12, Murillo13, Sakai14, Maret20}.
Dust grains coagulate and grow in the protostellar disk to at least mm-size and these grains are highly settled to a geometrically thin midplane \citep{Pinte16}.  During this stage the gas is still coupled to the dust and the budgets of highly volatile ices might be altered. Grain growth should enhance the penetration of UV photons, which are mediated by the small sub-micron grains in the system.
Overall, the presence of species such as CH$^+$ provides hints that UV photons are present in protostellar envelopes and irradiate outflow cavity walls \citep{Kristensen13, Benz16}.  In principle, this could increase the potential for self-shielding to produce O and N isotopic enrichments in the young disk as suggested by \citet{Lyons05} and \citet{Visser18}.   It is not clear that the UV penetrates to layers deep enough such that the products of UV photochemistry can be incorporated into forming ices to be carried down to the dust-rich protostellar disk midplane.

\subsection{What aspects of cometary material trace star and disk formation?}

\begin{figure*}[ht!]
\begin{center}
\includegraphics[width=10cm]{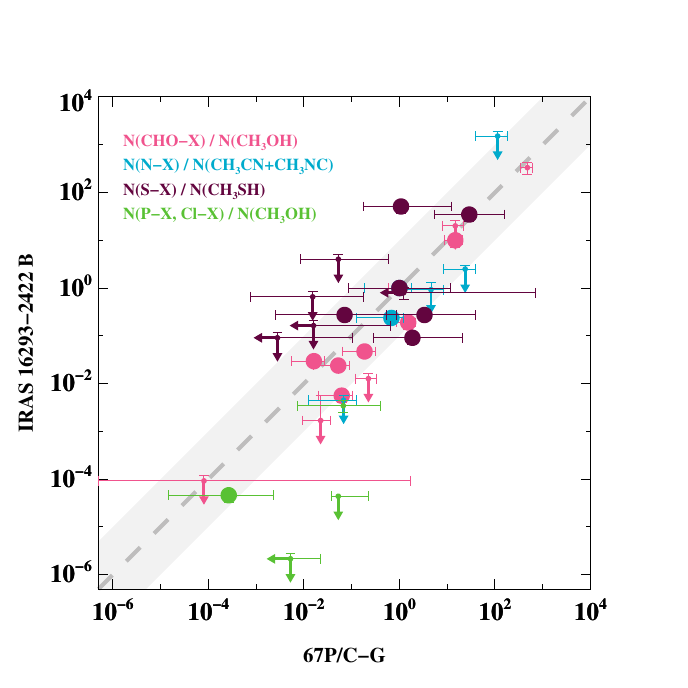}
\caption{The abundance of CHO-, N-, S-, P- and Cl-bearing molecules
relative to species specified in the top left corner (which include methanol, methyl cyanide, methyl isocyanide, and methyl mercaptan). The values along the ordinate were observed with ALMA towards an offset position near IRAS~16293-2422~B ($0.5\arcsec$ to the SW from the source). These observations are probing volatiles on disk scales in this young low-mass protostar. The values along the abscissa were measured with Rosetta-ROSINA in the coma of comet 67P/Churyumov–Gerasimenko. Each chemical family has a unique color. The shaded region corresponds to an order of magnitude scatter about the one-to-one linear correlation of the two date sets. Smaller sized data points pertain to an upper limit or an estimate of sorts either for the protostar, or the comet, or both. A significant correlation is seen between cometary and protostellar volatile abundances, suggesting preservation of prestellar and protostellar volatiles into cometary bodies upon some degree of chemical alteration. All details are available in the original source: \citet{Drozdovskaya2021}.}
\label{fig:67P_IRAS16293B}
\end{center}
\end{figure*}

\subsubsection{CAIs and crystalline silicates}

Analysis of the Stardust samples established that a significant fraction of cometary solids is the result either of high temperature processing in the disk (CAIs, chondrules) or low (370-420 K) temperature aqueous alteration in the Wild 2 comet, demonstrating that cometary solids are not entirely composed of preserved interstellar dust as originally thought \citep{Brownlee2014}. They also showed that the differences between cometary solids and carbonaceous chondrites is comparable to the differences observed within the carbonaceous chondrites themselves, changing our perspective on possible cometary meteorites.    The high temperature materials must be products of processes operating in the early protostellar disk. Generally, they are assumed to have formed relatively close to the Sun and could have been carried outwards by an initially hot ($>$1000~K) compact but expanding disk \citep{Ciesla15}.   This initially hot material will have had little D-enrichment and as it cooled would have made water ice with the cosmic D/H ratio ($\sim$10$^{-5}$).  To account for elevated cometary D/H ($\ge10^{-4}$ in H$_2$O) ratios would have required mixing of this inner solar system water with pristine material provided by infall from the cold (T $\sim 10-20$~K) core.

Comets are known to have crystalline silicates \citep{Crovisier97} intermixed with amorphous grains that must have formed at lower temperature \citep{Hanner99}. Since most interstellar silicates are amorphous \citep{Kemper04}, this means that these grains must have crystallized in the protostellar or the protoplanetary disk.  Thermal annealing of initially amorphous grains in the inner warmer regions of the disk combined with radial mixing is suggested as a potential process \citep[][and references therein]{Henning10}, along with shocks \citep{Harker02}.  Annealing would have been more strongly active in the younger warmer protostellar disk, and astronomical observations of young disks suggest that crystalline silicate emission comes predominantly from the inner regions the disks \citep{vanBoekel04, Bouwman08, Watson09}.

\noindent {\it Summary of Potential Links to Birth Environment:}   Both CAI's and crystalline silicates likely formed in the disk and, thus, are not linked to the birth environment.

\subsubsection{Overall chemical inventory}

There is an intriguing overlap between the volatile chemical composition of comets and young star-forming regions \citep{GreenbergLi1999c, EhrenfreundCharnley2000, MummaCharnley2011}. A young protostar heats up its surrounding environment (envelope and inner disk) resulting in a region where most volatiles are thermally desorbed into the gas, which is called the hot corino (or hot core in the case of high-mass protostars). The chemical inventories of hot corinos are a unique window on the complete volatile inventories in star-forming regions as volatile molecules are no longer hidden from observations in the ices. Ground-based observations of the Oort cloud comet C/1995 O1 (Hale-Bopp) showed a strong similarity between Hale-Bopp's abundances of CHO- and N-bearing molecules and those observed on large envelope- and cloud-scales (thousands of au; \citealt{Bockelee-Morvan2000}). These similarities were confirmed based on in situ mass spectrometry measurements by the Rosetta spacecraft at the Jupiter-family comet 67P/Churyumov-Gerasimenko, and on small disk-scales (tens of au) probed with ALMA (Figure~\ref{fig:67P_IRAS16293B}, \citealt{Drozdovskaya2019}, and recently extended to S-bearing molecules). 

\noindent {\it Summary of Potential Links to Birth Environment:} If other comets have chemical inventories that match those of C/1995 O1 (Hale-Bopp) and 67P/Churyumov-Gerasimenko, such similarities imply that the volatile chemical inventory of comets is to a degree set in the prestellar and protostellar stages, because the same molecules are found in both contexts with comparable relative ratios.    More work needs to be done to explore the differences between ice abundances in clustered vs isolated environments, by, e.g. JWST.

\begin{figure*}[ht!]
\begin{center}
\includegraphics[width=10cm]{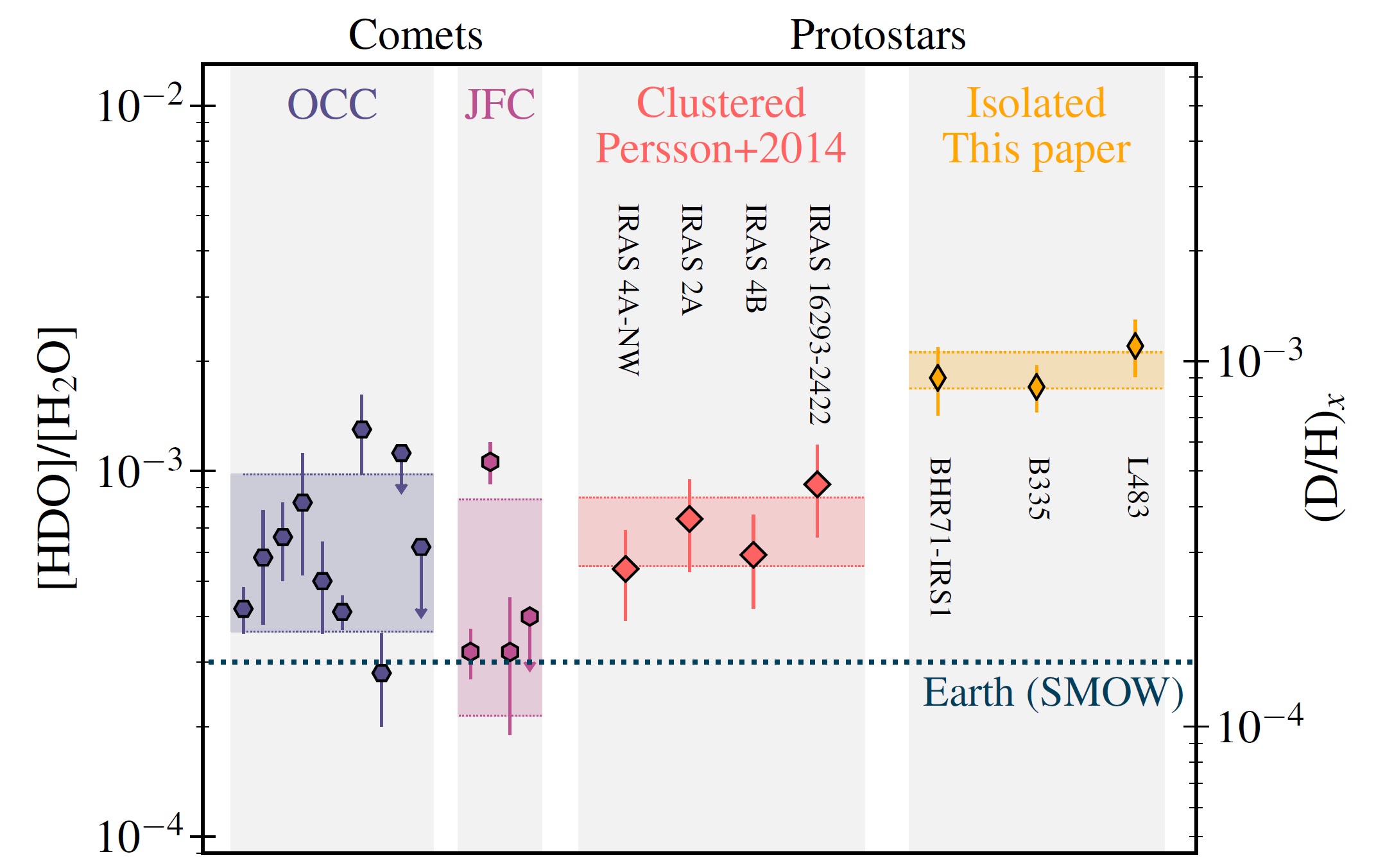}
\caption{The HDO/H$_{2}$O ratio along the left ordinate and the D/H ratio along the right ordinate for Oort Cloud Comets (OCC), Jupiter-Family Comets (JFC), and low-mass Class 0 protostars in clustered and isolated environments. Error bars are $1\sigma$ uncertainties. The colored regions are the standard deviations per category. In this publication, sources were classified as isolated if they are not associated with any known cloud complexes. The clustered regions pertain to star-forming regions with several low-mass protostars therein. Isolated sources display significantly higher levels of deuteration than clustered environments and solar system comets, suggesting that our Sun was born in a cluster. Full details are available in the original source: \citet{Jensen2019}.}
\label{fig:Clustered_Isolated_DH_Jensen}
\end{center}
\end{figure*}

\subsubsection{Isotopic ratios}
Isotopic ratios in volatiles also support the critical nature of early phases of star formation for cometary volatiles \citep{Bockelee-Morvan2015a}. Mass spectrometry measurement from the Rosetta mission to comet 67P/Churyumov–Gerasimenko on the ratio of D$_{2}$O/HDO to HDO/H$_{2}$O is much higher ($17$) than the statistically expected value of $0.25$ \citep{Altwegg2017a}. Overall, mono-deuterated water has a D/H ratio that is more than two times higher than the Vienna Standard Mean Ocean Water (VSMOW) value and in closer agreement with the elevated deuteration seen in star-forming regions \citep{Altwegg2019}.  
Observations carried out at high spatial resolution with ALMA suggest that the water D/H ratio can differ between clustered and isolated protostars (Figure~\ref{fig:Clustered_Isolated_DH_Jensen}). Water in isolated regions has D/H ratios that are a factor of $2-4$ higher than in clustered regions, which may be a result of either colder temperatures of the innate molecular cloud or longer collapse timescales in the isolated environments \citep{Jensen2019}. The effect of temperature on the overall deuterium chemistry is well documented via observations and theory \citep[see e.g.,][]{Punanova16}.  It should not be surprising that this could be reflected within ices that form as the result of this temperature dependent process.

Recently, the O isotopic ratio of water \citep{Schroeder19} and other species \citep{Altwegg20} have been found to be enhanced in $^{17}$O and $^{18}$O, relative to $^{16}$O and solar, in comet 67P.   For water, this effect is modest - $^{18}$O is enriched by 19$\pm$9\% ($\delta^{18}$O$_{VSMOW}$=121$\pm$89 \permil) and $^{17}$O by 28$\pm$10\% ($\delta^{17}$O$_{VSMOW}$=206$\pm$94 \permil) (1$\sigma$ errors). A similarly $^{16}$O-poor isotopic signature has been inferred for the initial Solar System water from meteoritic materials \citep{RN3860}. This signature could have originated in the young protostellar disk or in the protoplanetary disk \citep{Lyons05}, or the protosolar molecular cloud \citep{RN3860, Lee:2008, RN7221}.

\noindent {\it Summary of Potential Links to Birth Environment:}  Given existing information, comets in our solar system show a closer agreement with clustered star-forming regions in the measured water ice D/H ratio.   The oxygen and nitrogen isotopic ratio also bear information on the birth environment; however, more work needs to understand the potential contribution of the disk.


\subsubsection{Sulfur as a unique tracer}

\begin{figure*}[ht!]
\begin{center}
\includegraphics[width=10cm]{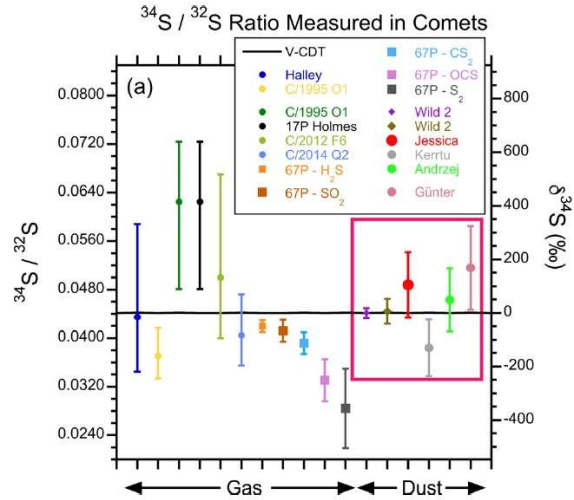}
\caption{The $^{34}$S/$^{32}$S isotopic ratios measured in several different comets. The figure includes measurements obtained during the Rosetta mission to comet 67P/Churyumov–Gerasimenko with the ROSINA instrument measuring volatile gases (squares with molecules specified) and with the COSIMA instrument measuring refractory dust grains (circles corresponding to named particles). Vienna-Canyon Diablo Troilite (V-CDT) is shown as a reference. The COSIMA measurements agree within errors with the V-DCT value and STARDUST values, while volatiles show a more significant dispersion. All details are available in the original source: \citet{Paquette2017}.}
\label{fig:67P_sulfur}
\end{center}
\end{figure*}

Sulfur is found in volatile cometary species (H$_{2}$S, SO$_{2}$, SO, OCS, H$_{2}$CS, CH$_{3}$SH, C$_{2}$H$_{6}$S, CS$_{2}$, CS; \citealt{Biver2016, Calmonte2016}) and in more refractory-like compounds (S$_{2}$, S$_{3}$), and even in atomic form \citep{Calmonte2016}. This implies that S is a unique element that may allow the volatile and refractory phases to be understood in relation to one another. The partitioning of elemental S between the two phases in comets has not been established yet \citep{Drozdovskaya2018, Altwegg2019}. From the ISM perspective, S is even more of a mystery. In the diffuse ISM, the cosmic abundance of S is accounted for fully in the gas phase \citep{LucasLiszt2002}. However, in prestellar cores and protostellar regions only a fraction of sulfur is observed in volatiles species \citep{Ruffle1999, Anderson2013}. It is thought that a large fraction of S may be hidden as H$_{2}$S ice; however, it has thus far not been identified via infrared observations \citep{Boogert1997, Boogert2015}. Alternatively, S may be predominantly found in more refractory phases \citep[e.g., S$_{2}$, S$_{3}$, FeS;][]{Kama19}. S$_{2}$ can be produced from H$_{2}$S ice upon ultraviolet (UV) irradiation \citep{GrimGreenberg1987}. Unfortunately, such species cannot be observed directly with remote facilities.

\noindent {\it Summary of Potential Links to Birth Environment:}  Sulfur isotopic ratios could be a promising way forward towards understanding the chemistry of S and how it gets incorporated into comets. Mass spectrometry measurements made by Rosetta on comet 67P/Churyumov–Gerasimenko allowed for the $^{34}$S/$^{32}$S ratio to be determined in volatiles (H$_{2}$S, OCS, CS$_{2}$; \citealt{Calmonte2017}) and in several dust particles \citep{Paquette2017}. The ratio in the refractories is consistent with the terrestrial value, but that in the volatiles is lower (Fig.~\ref{fig:67P_sulfur}). The isotopic ratio could potentially be used to trace the amount of S harbored in the refractory phase and unlock the potential of this abundant element as a tracer of origins.

\section{The incomplete picture of our origins}

 \subsection{A broader context of cometary}
 
For a long time, comets have been only looked at from a distance.  Their differences with asteroids were obvious since, by construction, comets were defined as active bodies. In contrast, asteroids are usually defined as activity-free objects. The focus on cometary ices, which show strong links with the ISM,  led to the idea that comets were primitive, unprocessed bodies. The identification of the highly processed Orgueil meteorites as a possible cometary meteorite, the discovery that Wild 2 cometary samples contained both high-temperature minerals and alteration phases, and the presence of active bodies in the outer belt \citep{Hsieh06}, changed this view and lead to the idea that there might exist a continuum between asteroids and comets. 

The recent discovery that carbonaceous chondrites and related meteorites probably came from the outer solar system supported the asteroid-comet continuum and that at least a significant fraction of cometary dust has been processed in the protoplanetary disk or in parent-bodies. Since there is a large variability within the different carbonaceous chondrite groups, it should also lead to the realization that there might be as much variability in comets as in asteroids, and that speaking of comets in general might be misleading. Finally, one should keep in mind that the phenomenological definition of comets as active bodies might hinder the fact that some so-called asteroids are also ice-rich and formed in the outer solar system, leading to a reconsideration of the notion of primitivity.

\subsection{Pinpointing t=0 in a Solar birth cluster}

The chemical composition of comets  is diverse and  complex. The similarity in composition  between the chemical inventory of cometary volatiles and gases in protoplanetary disk-forming regions in the vicinity of protostars suggests partial inheritance of materials. These similarities have now been demonstrated for more than one comet and are strengthened by improvements in the quality of the available data. Deuteration of cometary molecules and the presence of highly volatile species at appreciable abundances in them strongly supports a pristine cometary nucleus, one that has never been fully heated to engender significant volatile loss. Moreover, the high levels of deuteration of cometary volatiles, and indeed in solar system water, can only be achieved in the darkest, coldest parts of the ISM - the prestellar cores.    The alternative would be that these ratios are created in the cold outer parts of the disk, but the lack of a strong gradient in the water D/H ratios in the solar system, as would be expected given the temperature gradient, and the need for ionization \citep{Cleeves14} argue against the outer disk \citep[but, see][]{Furuya15}. This {\em may} be pinpointing the $t=0$ point of cometary volatile ice formation.

The D/H ratio is a key fingerprint.  The origin of O and N isotopic anomalies in meteorites, which appear to be present in cometary volatiles \citep{Schroeder19, Altwegg20}, could also represent further important clues. Self-shielding will be active during molecular cloud and dense core formation as these effects are widely observed in the ISM.  Models of the origin of heavy O isotope enrichment in interstellar ices within the molecular cloud require the birth of the Sun in a cluster \citep{Lee:2008}, in agreement with dynamical evidence (\S 3.1), and radionuclides (\S 2.3).  This is now in agreement with the water D/H ratio.  Regions associated with star cluster birth have (D/H)$_{\rm H_2O}$ $\sim 10^{-4}$, comparable to the solar system, with higher ratios measured in more isolated systems \citep{Jensen2019}.  Thus, isotopic enrichment may be a key new piece in support of the Sun's birth in a clustered environment. However, at present, origin within the disk, e.g., \citet[][]{Lyons05}, cannot be ruled out.  Clearly, more (and higher precision), measurements of O, N, S, and H isotopic ratios in comets are needed.  Further, more detailed chemical/dynamical models that encompass all isotopic systems need to be developed.

\subsection{Looking forward}

There remain significant open questions.  First, are the high temperature components mixed outward - prior to the main phase of cometary assembly - to be combined with the grains with ice coatings from earlier phases or are they somehow created in the comet-forming zone. \citet{Bergner21} points out that radial drift provides a strong supply term of primordial material to the inner tens of au. Thus, inward supply of solids is clear, but supply via outwards movement via, e.g., viscous spreading, is unsettled.  Analysis of solar system samples provide information, with the support of models, of outward movement of material \citep{Yang13, Desch18, Nanne19}.  Astronomically, evidence is less clear \citep{Najita18, Trapman22} and which mechanism (spreading or drift) dominates needs to be understood.    The carbonaceous refractory components are another important aspect with a new burst of information from Rosetta.   This component carries the bulk of C and N within cometary nuclei.  However, we know very little about its origin and need to develop remote sensing techniques to probe commonalities of this phase beyond the detailed in situ studies of 1P/Halley and 67P. The work of \citet{Woodward21} provides a recent example of how to explore this and \citet{Lisse20} summarizes the Spitzer mission legacy, which will be significantly expanded upon by the contributions of the James Webb Space Telescope.    There is some dispersion in the bulk cometary carbon content with potentially two Sun-grazing comets having bulk carbon consistent with chondrites \citep{Ciaravella10, McCauley13}.   This is significantly below 1P/Halley and 67P \citep{Bergin15, Rubin19} and hints at relatively unprobed diversity in the bulk content of cometary bodies.

More broadly the D/H ratio of water vapor is known with high precision in a handful of comets and whether there is a traceable link between D/H ratio (or other isotopic ratios) to, e.g., radius of formation, appears to be difficult.  It may be the case that radial drift wipes out any such signature, but more models, combined with a statistically significant sample observed with high precision is needed.

\vskip .5in
\noindent \textbf{Acknowledgments.} \\

E.A.B. acknowledges support from NSF Grant\#1907653 and NASA Exoplanets Research Program, grant 80NSSC20K0259 and Emerging Worlds Program, grant 80NSSC20K0333. 
M.N.D. acknowledges support of the Swiss National Science Foundation (SNSF) Ambizione grant no. 180079, the Center for Space and Habitability (CSH) Fellowship, and the IAU Gruber Foundation Fellowship. M.N.D. also acknowledges beneficial discussions held with the international team \#461 ``Provenances of our solar system's Relics'' (team leaders Maria N. Drozdovskaya and Cyrielle Opitom) at the International Space Science Institute, Bern, Switzerland.

\bibliographystyle{sss-full.bst}
\bibliography{z-refs.bib}
\end{document}